\documentclass[10pt,aps,pre,twocolumn,superscriptaddress,amsmath,amssymb]{revtex4-1}
\usepackage[utf8x]{inputenc}
\usepackage[english]{babel}
\usepackage{graphicx}
\usepackage[usenames,dvipsnames,svgnames,table]{xcolor}
\usepackage{hyperref}
\usepackage[caption=false]{subfig}
\usepackage{flafter}

\definecolor{col1}{rgb}{1,0.5,0}
\definecolor{col2}{rgb}{1,0.5,0}
\definecolor{col3}{rgb}{1,0.5,0}
\definecolor{col4}{rgb}{1,0.5,0}

\definecolor{col_lc}{HTML}{77ac30}
\definecolor{col_rc}{HTML}{0060ad}
\definecolor{col_l}{HTML}{888A85}
\definecolor{col_c}{HTML}{dd181f}

\usepackage{bm}

\usepackage{textcomp}
\newcommand{\plusovercross}{\rlap{+}{\texttimes}}

\begin{document}

\title{On the phase diagram of Mackay icosahedra}

\author{Marko Mravlak}
\affiliation{Physics and Materials Science Research Unit, Université du Luxembourg, L-1511 Luxembourg, Luxembourg}
\author{Tanja Schilling}
\affiliation{Physikalisches Institut, Albert-Ludwigs Universität Freiburg, D-79104 Freiburg, Germany}

\date[Date: ]{\today}

\begin{abstract}
Using Monte Carlo and molecular dynamics simulations, we investigate the equilibrium phase behavior of a monodisperse system of Mackay icosahedra.
We define the icosahedra as polyatomic molecules composed of a set of Lennard-Jones subparticles arranged on the surface of the Mackay icosahedron.
The phase diagram contains a fluid phase, a crystalline phase and a rotator phase.
We find that the attractive icosahedral molecules behave similar to hard geometric icosahedra for which the densest lattice packing and the rotator crystal phase have been identified before.
We show that both phases form under attractive interactions as well.
When heating the system from the dense crystal packing, there is first a transition to the rotator crystal and then another to a fluid phase.
\end{abstract}


\maketitle

\section{Introduction}

When a small number of atoms or molecules are forced together they tend to form locally densely packed clusters, in which a central atom is covered by 12 nearest neighbors located at the vertices of an icosahedron.
Since these structures posses a five fold symmetry, which is a forbidden crystal symmetry, they can not extend to large length scales. Instead they are often seen as favored, persisting local structures in supercooled liquids and glass-forming substances~\cite{Frank1952,Steinhardt1983}.

\begin{figure}
  \centering
  \includegraphics[width=\columnwidth]{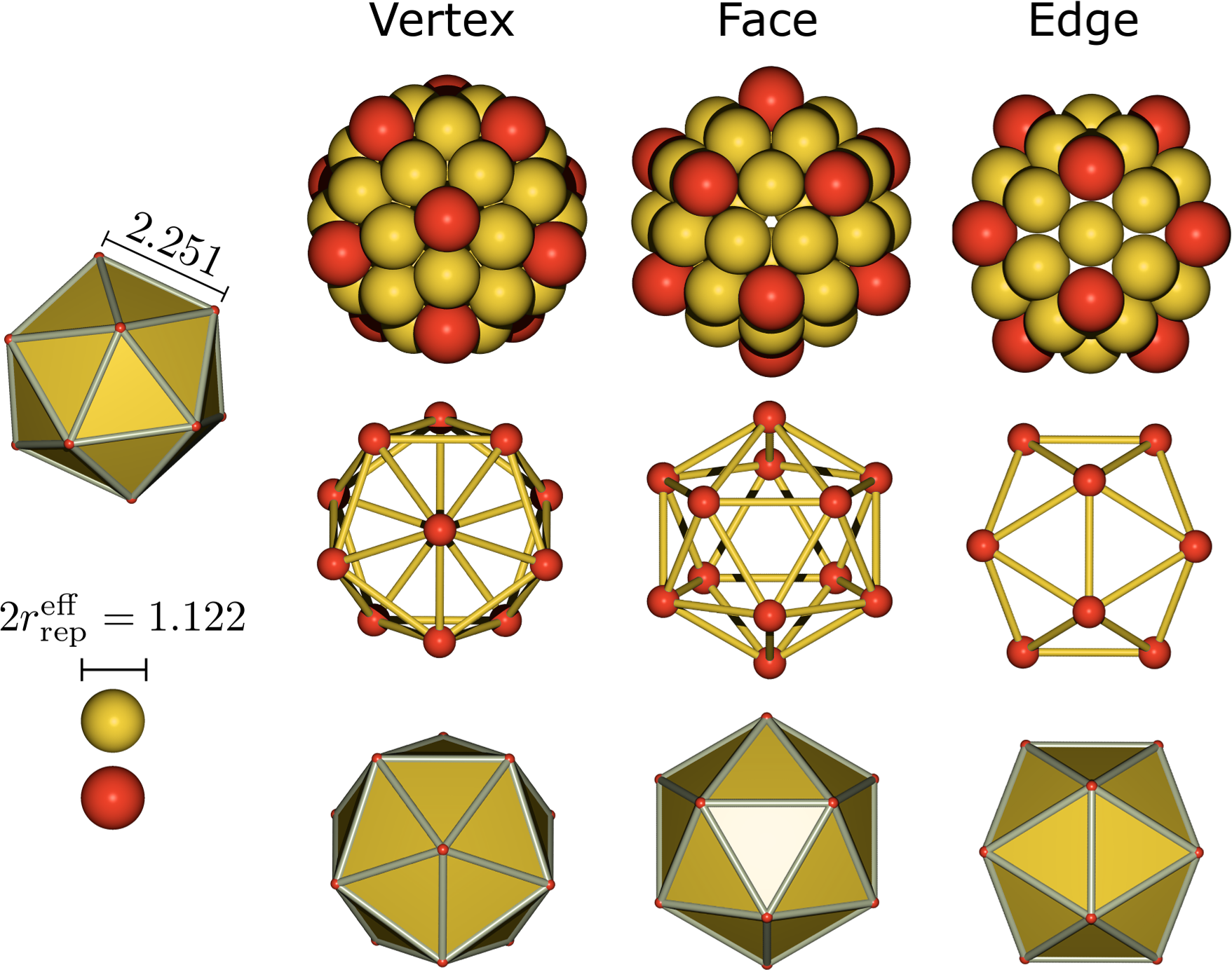}
  \caption{The Mackay icosahedron with two complete shells is composed of 55 subparticles of which only the 42 centers on the surface are considered for the model used in our simulations. The icosahedral molecule is shown in three different projections (vertex, face, edge on) and three visualization models (space filling, ball and stick, polyhedron) alongside with dimensions given in units of $\sigma$. Dark color denotes particles sitting at the 12 vertices of the shell. The radius of the spheres is the effective (repulsive) LJ radius $r_{\rm rep}^{\rm eff}=2^{1/6}\sigma/2$. To ease visualization of the model particles we present them as regular polyhedra with flat faces, round edges and vertices.
  }
  \label{fig:model}
\end{figure}

Mackay generalized this construction to larger multiply twinned superlattices of densely packed spherical subparticles with icosahedral symmetry~\cite{Mackay1962}. These are composed of 20 slightly deformed close-packed tetrahedra which are merged together in such a way that their adjacent planes form twinning crystal domains while their outer planes constitute the 20 faces of an icosahedron (Fig.~\ref{fig:model}).
If the number of constituent particles is below $\sim10^3$, Mackay icosahedra have been shown to globally minimize the Lennard-Jones potential energy surfaces~\cite{Wales2013}, to result from enthalpy-driven assembly of colloids~\cite{Rupich2009} and to maximize the entropy in spherical confinement of hard spheres~\cite{DeNijs2015}.
They are favored local structures in diverse systems such as noble gas atoms and molecules~\cite{Farges1986}, gold nanoparticles~\cite{Lacava2012} and clusters of metal atoms~\cite{Kuo2002}. 
They are thus ubiquitous in nature and constitute an important class of structures.

Interesting from a technological point of view, are in particular the clusters formed by confined colloids, as they could be used as building blocks in hierarchically structured materials (so called "meta-materials"). Confinement of spherical gold nanoparticles inside spherical surfaces of emulsion droplets leads to an assembly of nanocrystals with superlattices corresponding to Mackay icosahedra~\cite{Lacava2012}.
The structures of these clusters keep the polyhedral character also for binary mixtures of particles with different attractions~\cite{Mravlak2016} while differences in atmospheric pressure can determine the formation of either a complex binary crystal super-lattice, core-shell, or Janus clusters in dispersions of nanoparticle with binary size distributions~\cite{Kister2016}.
Mackay icosahedra have also been observed for assemblies of nanometer- and micrometer-sized hard, spherical colloids in spherical confinement~\cite{DeNijs2015}.
As such structures have been synthesized, we now aim to use them for bottom up design of functional materials. To determine the properties of these materials an understanding of their equilibrium phase behavior is a necessary prerequisite. We thus present here a study of the phase diagram of a suspension of Mackay icosahedra as they would result for instance from a confined colloidal aggregation experiment.

The infinite pressure and temperature boundary of the phase diagram for geometric icosahedra is already known (at infinite temperature the attractive part of the potential becomes irrelevant, thus the interaction becomes a pure hard-core repulsion). To find the optimal arrangements of congruent objects that do not tile space and their associated maximal density is an ancient mathematical challenge that remains unsolved for all non-trivially shaped objects except for the sphere - the densest packing of which had been conjectured already by Kepler and has only been proven about a decade ago~\cite{Hales2005}.
A more recent conjecture suggests that the densest packings of centrally symmetric Platonic and Archimedean solids are their corresponding optimal Bravais lattice packings~\cite{Torquato2009}.
For hard icosahedra (and several other regular polyhedra) the densest lattice packing (DLP) has been determined using different numerical optimization algorithms~\cite{Betke2000,Conway2006,Torquato2009}.
The putative optimal arrangement is a locally jammed packing occupying a volume fraction of 0.8363574 where each icosahedron contacts 12 neighbors and is represented by a Bravais lattice with a triclinic unit cell where each point contains a single uniformly oriented body (see Fig.~\ref{fig:snaps}).
Another work has shown that a dense fluid of hard icosahedra assembles into a close-packed (FCC or HCP) rotator crystal where particles are allowed to freely rotate about their lattice positions~\cite{Damasceno2012}.
Polyhedra with small asphericity and high rotational symmetry are expected to form such mesophases in between the disordered and crystalline states~\cite{Agarwal2011}.
Packing small numbers of hard icosahedra in a hard spherical container results in clusters that resemble or match sphere clusters (optimal sphere codes) despite significant faceting of these objects~\cite{Teich2016}.

The bulk behavior of Mackay icosahedra has not yet been studied systematically although such structures are ubiquitous in nature and they could have important implications for materials science~\cite{Lu2013,Nagel2017}.
In the present work we explore the equilibrium phase behavior at finite temperatures and pressures using Monte Carlo (MC) as well as Molecular Dynamics (MD) simulations.
We chose to study the phase behavior for icosahedra each composed of 55 particles arranged in a Mackay fashion as their size is small enough for the energy computation to be feasible, while they still capture the geometrical features of icosahedra and allows for a description of energetic attributes of such objects.
Similar models have been employed to study bulk behavior of hard and attractive polybead cuboids and binary polyhedra~\cite{John2005,Khadilkar2012}.
We show that the predicted DLP packing and rotator phases for hard icosahedra are also stable at finite temperatures.
At very low temperatures, we did not reach equilibrium, but rather observed a kinetically arrested phase, the structure of which depends on the initial condition and the simulated dynamics.

\section{Simulation methods}

The Mackay icosahedron shown in Fig.~\ref{fig:model} is composed of 55 sub-particles arranged in two complete shells around its center.
This quasi-spherical arrangement is a particularly stable motif in the minimal energy Lennard-Jones diagrams for binary systems of particles in a wide range of different attraction ratios. The coordinates of sub-particles were obtained from a database of minimal energy Lennard-Jones clusters~\cite{Mravlak2016}.
In our simulations we define an icosahedral molecule as a rigid arrangement of the 42 sub-particles from the outer shell of the energy-minimized Mackay icosahedron. The innermost sub-particles lie on the surface of the inscribed sphere that is considered to be impenetrable.
Some of the geometrical properties of this icosahedral molecule are given in Table~\ref{tab:1} where we see that the volume of the icosahedron fills only $58.2\%$ of the circumscribed sphere.
Due to this significant deviation from a spherical shape entropic effects are expected to play an important role in systems of such molecules.
(We estimated the volume of the molecule with a Monte Carlo integration using an effective repulsive radius of sub-particles, $r_{\rm rep}^{\rm eff}=2^{1/6}\sigma/2$, where $\sigma$ is the usual Lennard-Jones potential range parameter, the inscribed impenetrable sphere is included in this volume.)

\begin{table}
  \centering
  \begin{tabular}{lr}
    \hline
    inradius $r_{\rm in}/\sigma$ & 1.835 \\
    circumradius $r_{\rm out}/\sigma$ & 2.141 \\
    effective circumradius $r_{\rm out}^{\rm eff}/\sigma$ & 2.702 \\
    effective volume $V^{\rm eff}/\sigma^3$ & 48.091 \\
    equivalent spherical radius $r_{\rm ic}^{\rm ES}=(V^{\rm eff})^{1/3}$ & 3.637 \\
    scaled exclusion volume $V^{\rm eff}/V_{\rm out}^{\rm eff}$ & 0.582\\
    \hline
  \end{tabular}
  \caption{Geometrical properties of the Mackay icosahedron model particle (Fig.~\ref{fig:model}) used in the simulations. $\sigma$ is the Lennard-Jones potential range parameter describing interactions between pairs of sub-particles. The effective radius and volume take into account the repulsive range of the Lennard-Jones potential, $r_{\rm rep}^{\rm eff}=2^{1/6}\sigma/2$.}
  \label{tab:1}
\end{table}

The interaction energy between two icosahedral molecules depends on the distance between their centers and on their mutual orientation.
The icosahedron is a nonlinear, rigid molecule whose configuration is given by a translational vector of its center $r=(x,y,z)$ and a quaternion $q=(a,b,c,d)$ specifying its orientation.
To calculate the positions of all sub-particles after the rotation we use a rotation matrix given by
\begin{equation}
  R=\left(
  \begin{matrix}
    a^2+b^2-c^2-d^2 & 2(bc-ad) & 2(bd+ac) \\
    2(bc+ad) & a^2-b^2+c^2-d^2 & 2(cd-ab) \\
    2(bd-ac) & 2(cd+ab) & a^2-b^2-c^2+d^2
  \end{matrix}
  \right)\;.
\end{equation}
We then calculate a sum of Lennard-Jones terms of all the pairs of sub-particles
\begin{equation}
U(r_1,r_2,q_1,q_2)=4\epsilon\sum\limits_{i=1}^{N}\sum\limits_{j\ne i}^{N}\left[\left(\frac{\sigma}{r_{ij}}\right)^{12}-\left(\frac{\sigma}{r_{ij}}\right)^{6}\right]\;,
\end{equation}
where $r_{ij}$ is the distance between sub-particles $i$ and $j$ and $N=42$.
We apply a cutoff of $2.5\,\sigma$ to the interaction between two sub-particles and thus exclude the contribution of the core particles.
Additionally a cutoff of $7.0\,\sigma$ is applied on the interaction between a pair of icosahedra.
The effective (repulsive) diameter of an icosahedron is in a range $d_{\rm rep}^{\rm eff}\in[4.7922-5.4046]\sigma$.

To simulate the system, we use two independent methods. The first is Monte Carlo simulation~\cite{Metropolis1953} in the isobaric-isothermal ensemble with a variable box size and periodic boundary conditions \cite{Wood1968,Wood1970}.
To release the eventual stresses in the orthogonal simulation box we additionally employ Parrinello-Rahman sampling of the variable box shape~\cite{Parrinello1980,Parrinello1981,Najafabadi1983,Yashonath1985,Filion2009}.
Second, in order to check for independence of our results from details of the simulation method, we apply molecular dynamics to integrate the system's trajectory in the isobaric-isothermal ensemble using the molecular simulator package LAMMPS~\cite{Plimpton1995}.
Here, we define icosahedra as independent rigid bodies and use a Nosé-Hoover thermostat with chains~\cite{Martyna1992} to ensure constant temperature and pressure in a triclinic simulation box.
We integrate the translational and rotational motion of rigid bodies by applying time reversible algorithms utilizing the quaternion representation~\cite{Kamberaj2005}.
In both simulation methods the deformations of the simulation box are limited to avoid unphysically deformed systems, especially at high temperatures.

We carried out both, MC and MD for 500 monodisperse icosahedra assembled from 21000 Lennard-Jones subparticles for a range of temperatures and pressures to explore the possible equilibrium phases. In the following, $T$ is given in units $\epsilon/k_B$ and $p$ in units $\epsilon/\sigma^3$.

To initialize the simulations, we placed the icosahedra either in a low density disordered fluid arrangement without any overlaps between them, or on one of the crystal and rotator crystal structures that spontaneously formed during initial MC runs.
Two million Monte Carlo cycles were then performed from several different initial conditions at each given pressure and temperature.
Displacement and rotational moves were used to sample configurations, where random orientations of the molecules are generated by sampling quaternions on the surface of the 4D unit sphere~\cite{Vesely1982}.
To facilitate the equilibration at low temperatures we introduced moves where the icosahedra can be displaced globally to any location in the simulation box and moves where two randomly chosen icosahedra are brought together or displaced out of the sphere of influence of each other which is defined by a diameter of $7.0\,\sigma$.
These moves are needed to break the energetically favorable face-to-face alignment at low temperatures whose energy amounts to $-6\epsilon$ in case of perfect alignment.
Details about these moves are given in the Appendix.
We monitored the system's energy, the virial pressure, the simulation box volume and tilt angles, the mean-square displacement, acceptance rates and step sizes. All data shown in this paper has been taken after these values saturated. (As we discuss in detail below, this does not imply that all structures we show here are thermodynamically stable.)

We use several structural descriptors to analyze the positional and orientational behavior.
The positional order is monitored by computing the radial distribution function.
Steinhardt bond orientational order parameters are used to quantify the bond network of nearest neighbors~\cite{Steinhardt1983, Wang2005}.
In both quantities the icosahedra are replaced by points lying in their centers (centroids).
To measure the amount of orientational order in a system of nonlinear molecules one needs to define a set of characteristic vectors attached to the model molecule. In the case of Mackay icosahedra we chose either the normals of the 20 triangles or the vectors pointing from the center to the 12 vertices on the surface of icosahedron, the results are similar for these choices.
The orientational pair correlation function (OPCF) can then be defined as~\cite{Chen2014}
\begin{equation}
 g_\text{opcf}(r)=\frac{\sum\limits_{i=1}^{N}\sum\limits_{j\ne i}^{N}\delta(r-r_{ij}) \left(\sin^2\alpha_{ij}^{F_1}+\sin^2\alpha_{ij}^{F_2}\right)/2}{\sum\limits_{i=1}^{N}\sum\limits_{j\ne i}^{N}\delta(r-r_{ij})}
\end{equation}
where $\alpha_{ij}^{F_1}$ and $\alpha_{ij}^{F_2}$ denote the first and the second minimal angles formed by any pair of the characteristic vectors associated to particles $i$ and $j$ such that $\alpha_{ij}^{F_1}\le\alpha_{ij}^{F_2}$.
The values of such a pair correlation function can get arbitrarily close to zero for highly orientationally ordered systems while its maximum values are limited by the largest values of the minimal angles which depend on the choice of characteristic vectors. (This definition might be counter-intuitive, as order corresponds to a small value. We use it nevertheless as defined in ref.~\cite{Chen2014} in order to allow for direct comparison to work by other authors.)

\section{Phase behavior of icosahedra}

\begin{figure}
  \centering
  \subfloat[$T=3$, $p=1.4$]{{\includegraphics[width=.50\columnwidth]{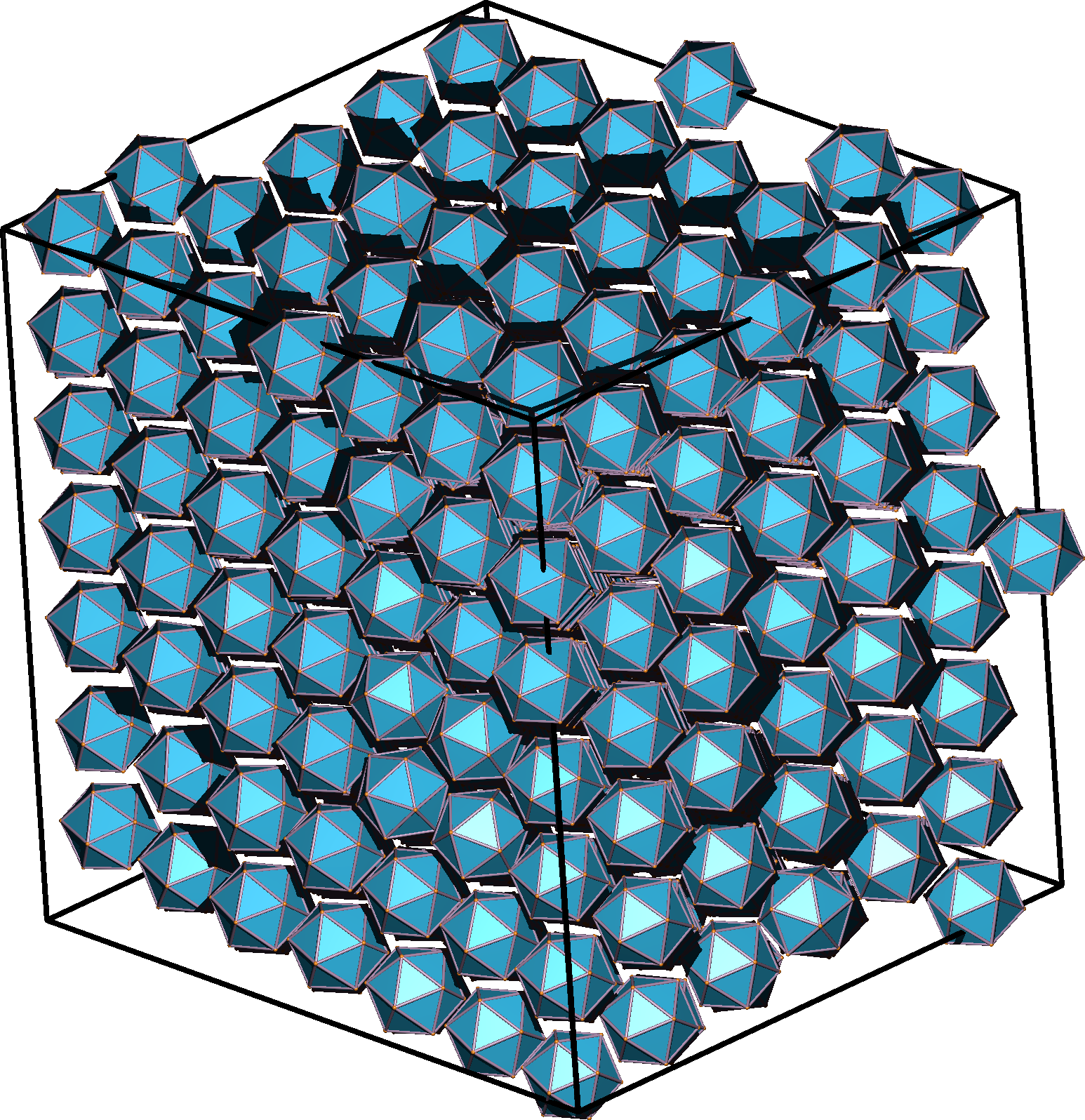}}\label{fig:snaps:a}}
  \subfloat[$T=4$, $p=1.4$]{{\includegraphics[width=.47\columnwidth]{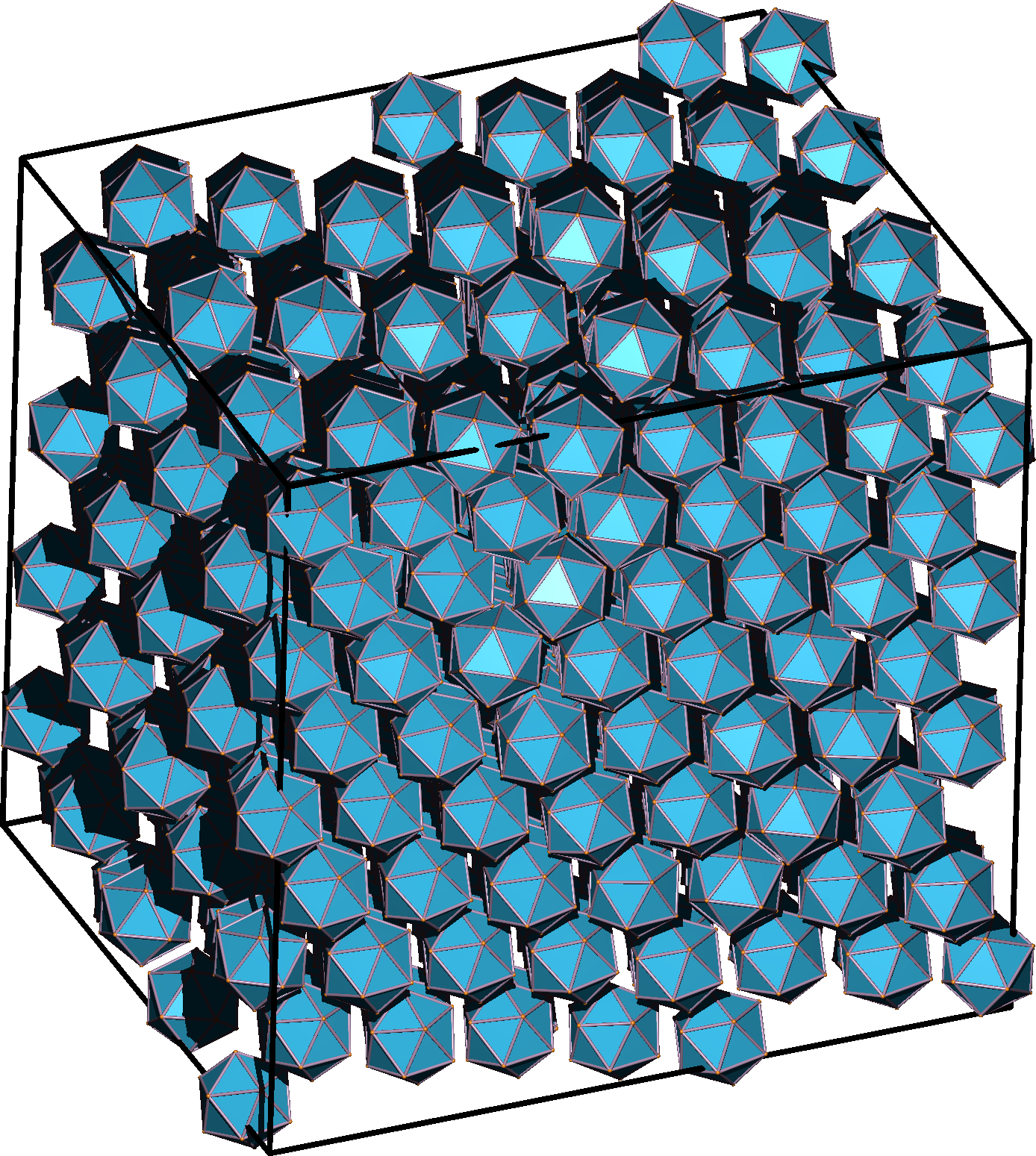}}\label{fig:snaps:b}}\\
  \subfloat[$T=6$, $p=1.4$]{{\includegraphics[width=.50\columnwidth]{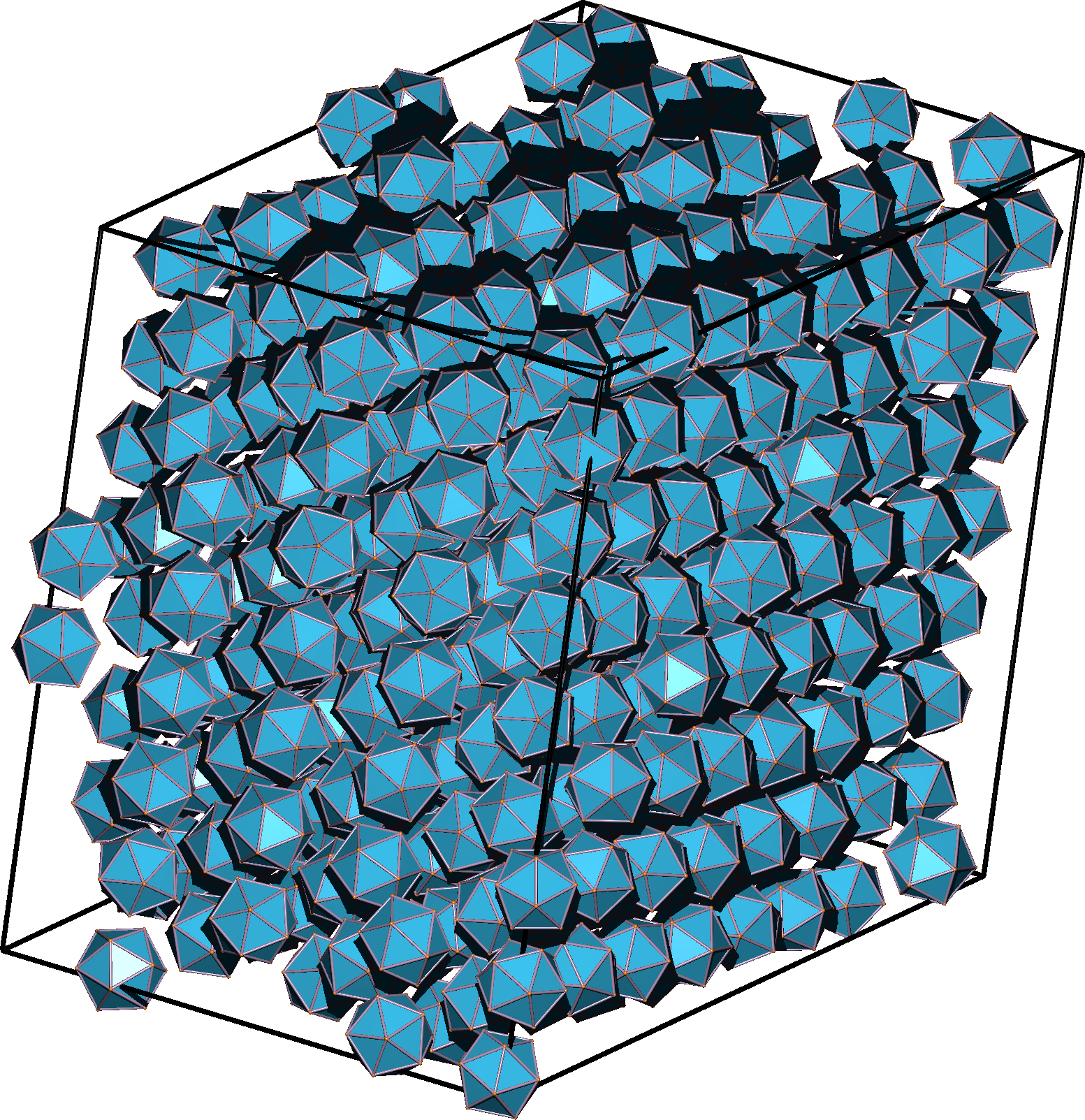}}\label{fig:snaps:c}}
  \subfloat[$T=13$, $p=1.0$]{{\includegraphics[width=.5\columnwidth]{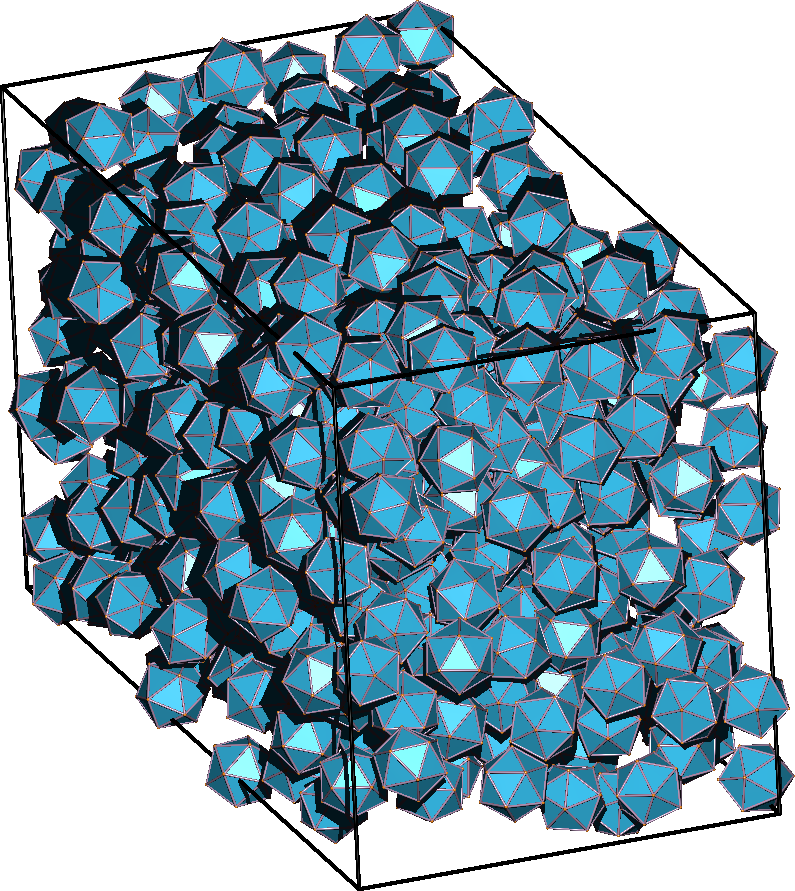}}\label{fig:snaps:d}}
  \caption{
  Snapshots of structures that formed in simulations with 500 icosahedra in a deformable parallelepiped with periodic boundary conditions.
  (a) A crystalline arrangement with a low potential energy and 14 nearest neighbors (LETC),
  (b) another orientationally aligned crystal phase, but here the icosahedra arranged on the conjectured DLP of hard icosahedra with a $7\,\%$ higher potential energy than (a).
  (c) At increased temperature a less orientationally ordered rotator crystal with FCC positional order formed.
  (d) At lower pressures and high temperatures a liquid phase with neither long range orientational nor positional order was obtained.
  The corresponding effective packing fractions, $\rho^{\rm eff}$, defined as the ratio of the volume taken by spheres with effective repulsive radius, are (a) $0.75$, (b) $0.73$, (c) $0.70$ and (d) $0.50$.
  }
  \label{fig:snaps}
\end{figure}

We observed five different structures which, once they had spontaneously formed out of the initial configurations, were persistent over the entire duration of the simulation: three competing crystal phases (a low potential energy triclinic crystal, abbreviated LETC in the following, shown in Fig.~\ref{fig:snaps:a}, DLP depicted in Fig.~\ref{fig:snaps:b} and a BCC phase that is not shown in Fig.~\ref{fig:snaps}), a rotator crystal phase shown in Fig.~\ref{fig:snaps:c} and a disordered liquid phase shown in Fig.~\ref{fig:snaps:d}.

\begin{figure}
  \includegraphics{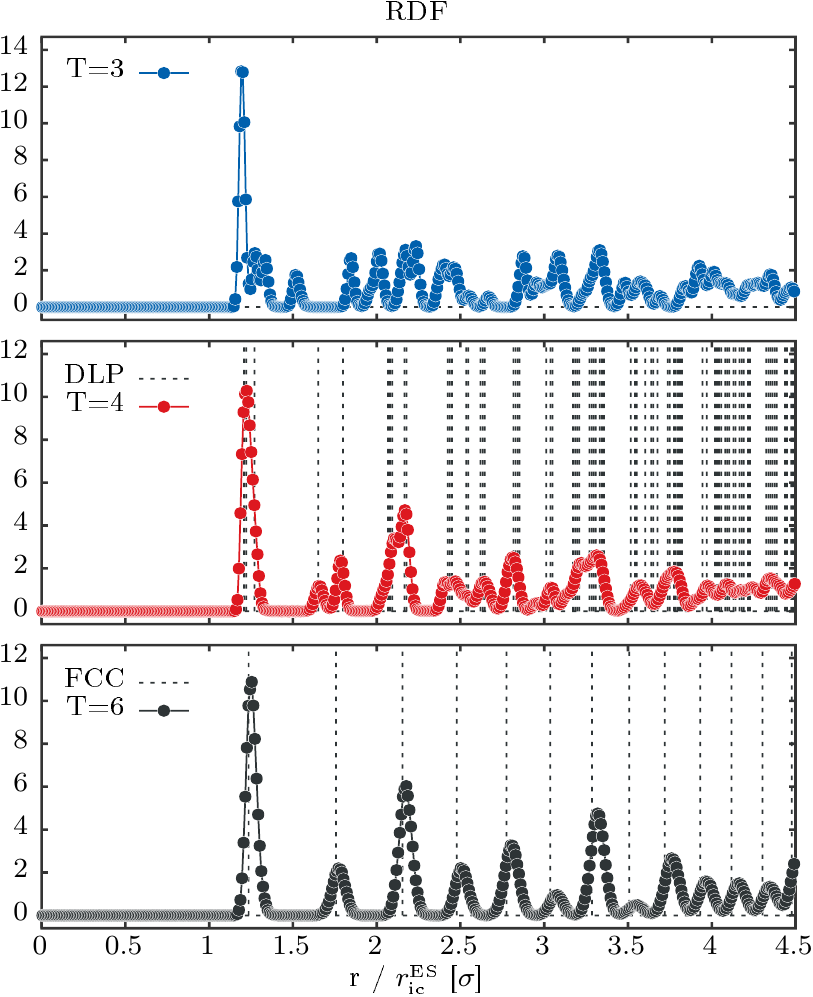}
  \caption{Radial distribution function for a system of 500 monodisperse icosahedra  at $p=1.4$ for three different temperatures. The pair distances are given in units of an equivalent spherical radius of icosahedra $r_{\rm ic}^{\rm ES}$. The dashed lines show the distances represented in a crystal with FCC or DLP lattice.}
  \label{fig:rdf}
\end{figure}

\begin{figure}
  \includegraphics{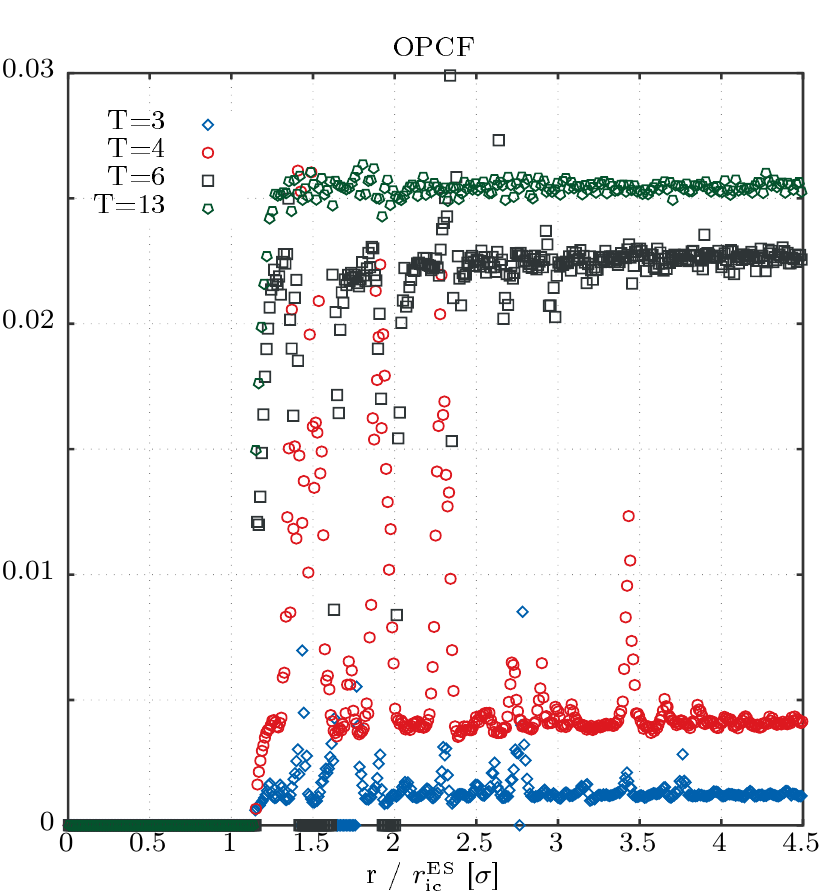}
  \caption{Orientational pair correlation function $g_\text{opcf}(r)$ for a system of 500 monodisperse icosahedra for four systems shown in Fig.~\ref{fig:snaps}.
  The orientational correlations were calculated using the normals of the 20 faces of an icosahedron as characteristic vectors.
  Low values of $g_\text{opcf}(r)$ represent aligned icosahedra at distance $r$ whereas large values characterize orientationally disordered pairs.
  The conferred systems display different degrees of orientational correlations with large degrees of long-range orientational order at $T=3$ and $T=4$, weak orientational order at $T=6$ and a fully orientationally disordered system at $T=13$.}
  \label{fig:opcf}
\end{figure}

The radial distribution functions in Fig.~\ref{fig:rdf} show the arrangement on a DLP and on a triclinic crystal lattice with 14 neighbors (LETC) at lower temperatures. At higher temperatures the system transforms into an FCC crystal lattice.
The corresponding orientational pair correlations were computed using the normals of the 20 faces of an icosahedron as characteristic vectors. We show them in Fig.~\ref{fig:opcf} for the four structures from Fig.~\ref{fig:snaps}.
Small values correspond to high mutual alignment while the strong oscillations are a consequence of regular particle positions as the OPCF is defined analogously to the RDF histogram but with different weighting factors.
The two crystalline systems at $T=3$ and $T=4$ have long-range orientational correlations as confirmed by low values of $g_\text{opcf}(r)$ and by visual inspection where we see that the icosahedra are almost entirely uniformly aligned.
In the remaining two structures at higher temperatures the icosahedra exhibit a lower degree of orientational order as confirmed by large values of $g_\text{opcf}(r)$.
The limiting value, which corresponds to a fully orientationally disordered system, depends on the choice of the characteristic vectors. In our case it is approximately 0.025.
With increase in temperature different average values of $g_\text{opcf}(r)$ are obtained indicating that the orientational order gradually decreases and that not all aligned or disordered phases exhibit the same level of orientational correlations.

When cooling the system from a completely disordered fluid at high temperatures, we observe the same trend at all the values of the pressure that we simulated.
At first, on decreasing the temperature, positional order appears in the system and a rotator crystal phase forms, while the orientational correlations are not yet established.
On further decreasing the temperature orientational order appears and the system forms a dense solid phase.
This confirms our expectations that the exact geometrical shape of an icosahedron is relevant at low temperatures, while at high temperatures the icosahedra behave like spheres with a hard core and a weak, short ranged attraction, and therefore tend to pack in the FCC lattice.
The increased importance of energetic interactions at low temperatures favors arrangements with larger numbers of pairwise bonds of sub-particles from neighboring icosahedra and thus lattices with lower densities but larger coordination numbers appear to compete with the DLP predicted for hard icosahedra. At very low temperatures the contribution of attractive energetic interactions between sub-particles becomes so important, that it prevents the system from reaching the equilibrium despite the MC moves that we used to break up face-to-face bonds.

\begin{figure}
  \centering
  \includegraphics{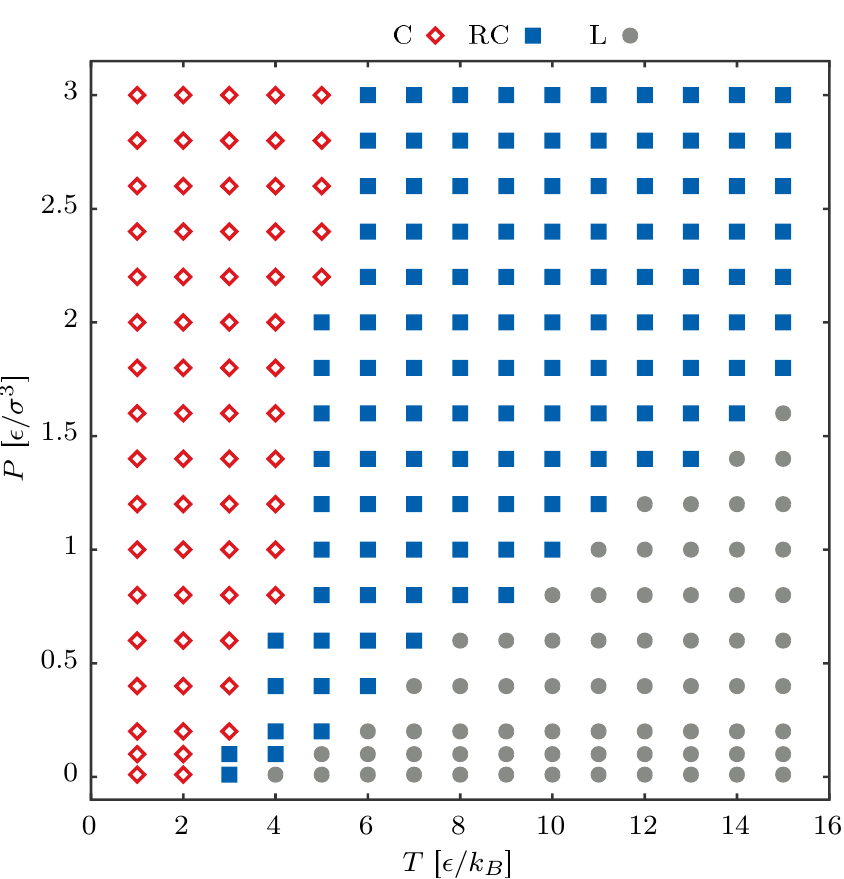}
  \caption{Scan of the phase diagram. The structures are classified into three categories, liquid (circles), rotator crystals with an FCC positional arrangement (squares) and a crystal region (diamonds) based the average value of the OPCF and the values of the average bond order parameters $Q_4$ and $Q_6$ as well as the comparison of peaks in RDF.}
  \label{fig:diag1}
\end{figure}

To scan the phase diagram we simulated at a number of different pressures and temperatures starting from several initial conditions.
The combined results are shown in Fig.~\ref{fig:diag1} where we labeled the phases as either crystal, rotator crystal with FCC positional order or liquid, according to a sequence of criteria explained in the appendix. The rotator crystal phase and the liquid phase are reached independently of the initial conditions and simulation method, thus we are confident that they are thermodynamically stable.

At low temperatures, simulations initialized in slightly different disordered fluid configurations evolved into significantly different solid structures.
Fig.~\ref{fig:diags1} shows results for three different initial configurations. The simulation runs either ended in a dense disordered packing or in one of several competing crystal lattices for the same phase point.
To further explore this we analyzed simulation runs from three additional initial conditions, the BCC and DLP lattices and an FCC rotator crystal, all of which had formed spontaneously in the previous runs.
The results are shown in Fig.~\ref{fig:diags2}. We observe that in the high density region at low temperatures the initial crystal structure has not changed for the initialized BCC and DLP lattices, while the initialized rotator crystal evolved into DLP except at the lowest temperatures. Often also the crystal phase with 14 neighbors formed.
The intermediate rotator crystal region remained unchanged for all initial conditions. We conclude that the high density disordered state we find at very low temperatures, is kinetically arrested and not thermodynamically stable. Whether the DLP arrangement (which is expected for hard icosahedra without energetic interactions), the BCC crystal or the other crystalline lattice, with the higher coordination number and lower potential energy, is thermodynamically most stable, needs to be decided by a future free energy calculation.

\section{Conclusions}

We have investigated the phase behavior of a monodisperse system of Mackay icosahedra composed from Lennard-Jones sub-particles. This type of cluster is a representative nanocrystal structure~\cite{Boles2016} that can be formed in experiments with agglomeration inside emulsion droplets~\cite{Lacava2012,DeNijs2015}.
Analysis of the positional and orientational pair correlation functions has shown a liquid phase, a rotationally disordered crystal with an FCC crystal lattice, and a dense crystalline phase. 

While the former two phases are thermodynamically stable, the structure of the latter depends sensitively on the simulation details: on changing the initial conditions, the simulations produced either a disordered dense packing, a uniformly oriented crystal phase with BCC ordering, the DLP of hard icosahedra, or another crystal lattice with a low potential energy and 14 nearest neighbors.
Simulations initialized in other structures than the DLP transformed into the DLP, but not the other way around. This indicates that the DLP is probably the equilibrium phase. However, there might be a range of low temperatures where the other crystalline phases are more stable.

While it takes a considerable amount of time to evaluate all the pair energies of the sub-particles within the cutoff range, the model also offers several interesting possibilities to generalize the components under study.
For example, interesting structures to study would be those where the vertices have modified attractions, describing functionalized interaction sites in nanocrystals which can be produced e.g.~by DNA-based ligands~\cite{Zhang2013} or the inclusion of magnetic sub-particles~\cite{Ye2016}.
This would enable us to simulate the assembly of the patchy building blocks in the form of binary icosahedral clusters with symmetric or Janus arrangements of interacting sites on the surface of clusters, which have been predicted before~\cite{Mravlak2016}.

\section{Acknowledgments}
This project has been financially supported by the National Research Fund (FNR) within the INTER-DFG project Agglo. Computer simulations presented in this paper were carried out using the HPC facilities of the University of Luxembourg.

\bibliographystyle{apsrev4-1}
\bibliography{refs}

\begin{thebibliography}{40}%
\makeatletter
\providecommand \@ifxundefined [1]{%
 \@ifx{#1\undefined}
}%
\providecommand \@ifnum [1]{%
 \ifnum #1\expandafter \@firstoftwo
 \else \expandafter \@secondoftwo
 \fi
}%
\providecommand \@ifx [1]{%
 \ifx #1\expandafter \@firstoftwo
 \else \expandafter \@secondoftwo
 \fi
}%
\providecommand \natexlab [1]{#1}%
\providecommand \enquote  [1]{``#1''}%
\providecommand \bibnamefont  [1]{#1}%
\providecommand \bibfnamefont [1]{#1}%
\providecommand \citenamefont [1]{#1}%
\providecommand \href@noop [0]{\@secondoftwo}%
\providecommand \href [0]{\begingroup \@sanitize@url \@href}%
\providecommand \@href[1]{\@@startlink{#1}\@@href}%
\providecommand \@@href[1]{\endgroup#1\@@endlink}%
\providecommand \@sanitize@url [0]{\catcode `\\12\catcode `\$12\catcode
  `\&12\catcode `\#12\catcode `\^12\catcode `\_12\catcode `\%12\relax}%
\providecommand \@@startlink[1]{}%
\providecommand \@@endlink[0]{}%
\providecommand \url  [0]{\begingroup\@sanitize@url \@url }%
\providecommand \@url [1]{\endgroup\@href {#1}{\urlprefix }}%
\providecommand \urlprefix  [0]{URL }%
\providecommand \Eprint [0]{\href }%
\providecommand \doibase [0]{http://dx.doi.org/}%
\providecommand \selectlanguage [0]{\@gobble}%
\providecommand \bibinfo  [0]{\@secondoftwo}%
\providecommand \bibfield  [0]{\@secondoftwo}%
\providecommand \translation [1]{[#1]}%
\providecommand \BibitemOpen [0]{}%
\providecommand \bibitemStop [0]{}%
\providecommand \bibitemNoStop [0]{.\EOS\space}%
\providecommand \EOS [0]{\spacefactor3000\relax}%
\providecommand \BibitemShut  [1]{\csname bibitem#1\endcsname}%
\let\auto@bib@innerbib\@empty
\bibitem [{\citenamefont {Frank}(1952)}]{Frank1952}%
  \BibitemOpen
  \bibfield  {author} {\bibinfo {author} {\bibfnamefont {F.~C.}\ \bibnamefont
  {Frank}},\ }\href {\doibase 10.1098/rspa.1952.0194} {\bibfield  {journal}
  {\bibinfo  {journal} {Proc. Royal Soc. A}\ }\textbf {\bibinfo {volume}
  {215}},\ \bibinfo {pages} {43} (\bibinfo {year} {1952})}\BibitemShut
  {NoStop}%
\bibitem [{\citenamefont {Steinhardt}\ \emph {et~al.}(1983)\citenamefont
  {Steinhardt}, \citenamefont {Nelson},\ and\ \citenamefont
  {Ronchetti}}]{Steinhardt1983}%
  \BibitemOpen
  \bibfield  {author} {\bibinfo {author} {\bibfnamefont {P.~J.}\ \bibnamefont
  {Steinhardt}}, \bibinfo {author} {\bibfnamefont {D.~R.}\ \bibnamefont
  {Nelson}}, \ and\ \bibinfo {author} {\bibfnamefont {M.}~\bibnamefont
  {Ronchetti}},\ }\href {\doibase 10.1103/PhysRevB.28.784} {\bibfield
  {journal} {\bibinfo  {journal} {Phys. Rev. B}\ }\textbf {\bibinfo {volume}
  {28}},\ \bibinfo {pages} {784} (\bibinfo {year} {1983})}\BibitemShut
  {NoStop}%
\bibitem [{\citenamefont {Mackay}(1962)}]{Mackay1962}%
  \BibitemOpen
  \bibfield  {author} {\bibinfo {author} {\bibfnamefont {A.~L.}\ \bibnamefont
  {Mackay}},\ }\href {\doibase 10.1107/S0365110X6200239X} {\bibfield  {journal}
  {\bibinfo  {journal} {Acta Crystallogr.}\ }\textbf {\bibinfo {volume} {15}},\
  \bibinfo {pages} {916} (\bibinfo {year} {1962})}\BibitemShut {NoStop}%
\bibitem [{\citenamefont {Wales}(2013)}]{Wales2013}%
  \BibitemOpen
  \bibfield  {author} {\bibinfo {author} {\bibfnamefont {D.~J.}\ \bibnamefont
  {Wales}},\ }\href {\doibase 10.1016/j.cplett.2013.07.066} {\bibfield
  {journal} {\bibinfo  {journal} {Chem. Phys. Lett.}\ }\textbf {\bibinfo
  {volume} {584}},\ \bibinfo {pages} {1} (\bibinfo {year} {2013})}\BibitemShut
  {NoStop}%
\bibitem [{\citenamefont {Rupich}\ \emph {et~al.}(2009)\citenamefont {Rupich},
  \citenamefont {Shevchenko}, \citenamefont {Bodnarchuk}, \citenamefont {Lee},\
  and\ \citenamefont {Talapin}}]{Rupich2009}%
  \BibitemOpen
  \bibfield  {author} {\bibinfo {author} {\bibfnamefont {S.~M.}\ \bibnamefont
  {Rupich}}, \bibinfo {author} {\bibfnamefont {E.~V.}\ \bibnamefont
  {Shevchenko}}, \bibinfo {author} {\bibfnamefont {M.~I.}\ \bibnamefont
  {Bodnarchuk}}, \bibinfo {author} {\bibfnamefont {B.}~\bibnamefont {Lee}}, \
  and\ \bibinfo {author} {\bibfnamefont {D.~V.}\ \bibnamefont {Talapin}},\
  }\href {\doibase 10.1021/ja9074425} {\bibfield  {journal} {\bibinfo
  {journal} {J. Am. Chem. Soc.}\ }\textbf {\bibinfo {volume} {132}},\ \bibinfo
  {pages} {289} (\bibinfo {year} {2009})}\BibitemShut {NoStop}%
\bibitem [{\citenamefont {de~Nijs}\ \emph {et~al.}(2015)\citenamefont
  {de~Nijs}, \citenamefont {Dussi}, \citenamefont {Smallenburg}, \citenamefont
  {Meeldijk}, \citenamefont {Groenendijk}, \citenamefont {Filion},
  \citenamefont {Imhof}, \citenamefont {van Blaaderen},\ and\ \citenamefont
  {Dijkstra}}]{DeNijs2015}%
  \BibitemOpen
  \bibfield  {author} {\bibinfo {author} {\bibfnamefont {B.}~\bibnamefont
  {de~Nijs}}, \bibinfo {author} {\bibfnamefont {S.}~\bibnamefont {Dussi}},
  \bibinfo {author} {\bibfnamefont {F.}~\bibnamefont {Smallenburg}}, \bibinfo
  {author} {\bibfnamefont {J.~D.}\ \bibnamefont {Meeldijk}}, \bibinfo {author}
  {\bibfnamefont {D.~J.}\ \bibnamefont {Groenendijk}}, \bibinfo {author}
  {\bibfnamefont {L.}~\bibnamefont {Filion}}, \bibinfo {author} {\bibfnamefont
  {A.}~\bibnamefont {Imhof}}, \bibinfo {author} {\bibfnamefont
  {A.}~\bibnamefont {van Blaaderen}}, \ and\ \bibinfo {author} {\bibfnamefont
  {M.}~\bibnamefont {Dijkstra}},\ }\href {\doibase 10.1038/nmat4072} {\bibfield
   {journal} {\bibinfo  {journal} {Nat. Mater.}\ }\textbf {\bibinfo {volume}
  {14}},\ \bibinfo {pages} {56} (\bibinfo {year} {2015})}\BibitemShut {NoStop}%
\bibitem [{\citenamefont {Farges}\ \emph {et~al.}(1986)\citenamefont {Farges},
  \citenamefont {De~Feraudy}, \citenamefont {Raoult},\ and\ \citenamefont
  {Torchet}}]{Farges1986}%
  \BibitemOpen
  \bibfield  {author} {\bibinfo {author} {\bibfnamefont {J.}~\bibnamefont
  {Farges}}, \bibinfo {author} {\bibfnamefont {M.}~\bibnamefont {De~Feraudy}},
  \bibinfo {author} {\bibfnamefont {B.}~\bibnamefont {Raoult}}, \ and\ \bibinfo
  {author} {\bibfnamefont {G.}~\bibnamefont {Torchet}},\ }\href {\doibase
  10.1063/1.450235} {\bibfield  {journal} {\bibinfo  {journal} {J. Chem.
  Phys.}\ }\textbf {\bibinfo {volume} {84}},\ \bibinfo {pages} {3491} (\bibinfo
  {year} {1986})}\BibitemShut {NoStop}%
\bibitem [{\citenamefont {Lacava}\ \emph {et~al.}(2012)\citenamefont {Lacava},
  \citenamefont {Born},\ and\ \citenamefont {Kraus}}]{Lacava2012}%
  \BibitemOpen
  \bibfield  {author} {\bibinfo {author} {\bibfnamefont {J.}~\bibnamefont
  {Lacava}}, \bibinfo {author} {\bibfnamefont {P.}~\bibnamefont {Born}}, \ and\
  \bibinfo {author} {\bibfnamefont {T.}~\bibnamefont {Kraus}},\ }\href
  {\doibase 10.1021/nl3013659} {\bibfield  {journal} {\bibinfo  {journal} {Nano
  Lett.}\ }\textbf {\bibinfo {volume} {12}},\ \bibinfo {pages} {3279} (\bibinfo
  {year} {2012})}\BibitemShut {NoStop}%
\bibitem [{\citenamefont {Kuo}(2002)}]{Kuo2002}%
  \BibitemOpen
  \bibfield  {author} {\bibinfo {author} {\bibfnamefont {K.~H.}\ \bibnamefont
  {Kuo}},\ }\href {\doibase 10.1023/A:1015847520094} {\bibfield  {journal}
  {\bibinfo  {journal} {Struct. Chem.}\ }\textbf {\bibinfo {volume} {13}},\
  \bibinfo {pages} {221} (\bibinfo {year} {2002})}\BibitemShut {NoStop}%
\bibitem [{\citenamefont {Mravlak}\ \emph {et~al.}(2016)\citenamefont
  {Mravlak}, \citenamefont {Kister}, \citenamefont {Kraus},\ and\ \citenamefont
  {Schilling}}]{Mravlak2016}%
  \BibitemOpen
  \bibfield  {author} {\bibinfo {author} {\bibfnamefont {M.}~\bibnamefont
  {Mravlak}}, \bibinfo {author} {\bibfnamefont {T.}~\bibnamefont {Kister}},
  \bibinfo {author} {\bibfnamefont {T.}~\bibnamefont {Kraus}}, \ and\ \bibinfo
  {author} {\bibfnamefont {T.}~\bibnamefont {Schilling}},\ }\href {\doibase
  10.1063/1.4954938} {\bibfield  {journal} {\bibinfo  {journal} {J. Chem.
  Phys.}\ }\textbf {\bibinfo {volume} {145}},\ \bibinfo {pages} {024302}
  (\bibinfo {year} {2016})}\BibitemShut {NoStop}%
\bibitem [{\citenamefont {Kister}\ \emph {et~al.}(2016)\citenamefont {Kister},
  \citenamefont {Mravlak}, \citenamefont {Schilling},\ and\ \citenamefont
  {Kraus}}]{Kister2016}%
  \BibitemOpen
  \bibfield  {author} {\bibinfo {author} {\bibfnamefont {T.}~\bibnamefont
  {Kister}}, \bibinfo {author} {\bibfnamefont {M.}~\bibnamefont {Mravlak}},
  \bibinfo {author} {\bibfnamefont {T.}~\bibnamefont {Schilling}}, \ and\
  \bibinfo {author} {\bibfnamefont {T.}~\bibnamefont {Kraus}},\ }\href
  {\doibase 10.1039/C6NR01940D} {\bibfield  {journal} {\bibinfo  {journal}
  {Nanoscale}\ }\textbf {\bibinfo {volume} {8}},\ \bibinfo {pages} {13377}
  (\bibinfo {year} {2016})}\BibitemShut {NoStop}%
\bibitem [{\citenamefont {Hales}(2005)}]{Hales2005}%
  \BibitemOpen
  \bibfield  {author} {\bibinfo {author} {\bibfnamefont {T.~C.}\ \bibnamefont
  {Hales}},\ }\href {\doibase 10.4007/annals.2005.162.1065} {\bibfield
  {journal} {\bibinfo  {journal} {Ann. Math.}\ }\textbf {\bibinfo {volume}
  {162}},\ \bibinfo {pages} {1065} (\bibinfo {year} {2005})}\BibitemShut
  {NoStop}%
\bibitem [{\citenamefont {Torquato}\ and\ \citenamefont
  {Jiao}(2009)}]{Torquato2009}%
  \BibitemOpen
  \bibfield  {author} {\bibinfo {author} {\bibfnamefont {S.}~\bibnamefont
  {Torquato}}\ and\ \bibinfo {author} {\bibfnamefont {Y.}~\bibnamefont
  {Jiao}},\ }\href {\doibase 10.1038/nature08239} {\bibfield  {journal}
  {\bibinfo  {journal} {Nature}\ }\textbf {\bibinfo {volume} {460}},\ \bibinfo
  {pages} {876} (\bibinfo {year} {2009})}\BibitemShut {NoStop}%
\bibitem [{\citenamefont {Betke}\ and\ \citenamefont {Henk}(2000)}]{Betke2000}%
  \BibitemOpen
  \bibfield  {author} {\bibinfo {author} {\bibfnamefont {U.}~\bibnamefont
  {Betke}}\ and\ \bibinfo {author} {\bibfnamefont {M.}~\bibnamefont {Henk}},\
  }\href {\doibase 10.1016/S0925-7721(00)00007-9} {\bibfield  {journal}
  {\bibinfo  {journal} {Comput. Geom.}\ }\textbf {\bibinfo {volume} {16}},\
  \bibinfo {pages} {157} (\bibinfo {year} {2000})}\BibitemShut {NoStop}%
\bibitem [{\citenamefont {Conway}\ and\ \citenamefont
  {Torquato}(2006)}]{Conway2006}%
  \BibitemOpen
  \bibfield  {author} {\bibinfo {author} {\bibfnamefont {J.~H.}\ \bibnamefont
  {Conway}}\ and\ \bibinfo {author} {\bibfnamefont {S.}~\bibnamefont
  {Torquato}},\ }\href {\doibase 10.1073/pnas.0601389103} {\bibfield  {journal}
  {\bibinfo  {journal} {Proc. Natl. Acad. Sci.}\ }\textbf {\bibinfo {volume}
  {103}},\ \bibinfo {pages} {10612} (\bibinfo {year} {2006})}\BibitemShut
  {NoStop}%
\bibitem [{\citenamefont {Damasceno}\ \emph {et~al.}(2012)\citenamefont
  {Damasceno}, \citenamefont {Engel},\ and\ \citenamefont
  {Glotzer}}]{Damasceno2012}%
  \BibitemOpen
  \bibfield  {author} {\bibinfo {author} {\bibfnamefont {P.~F.}\ \bibnamefont
  {Damasceno}}, \bibinfo {author} {\bibfnamefont {M.}~\bibnamefont {Engel}}, \
  and\ \bibinfo {author} {\bibfnamefont {S.~C.}\ \bibnamefont {Glotzer}},\
  }\href {\doibase 10.1126/science.1220869} {\bibfield  {journal} {\bibinfo
  {journal} {Science}\ }\textbf {\bibinfo {volume} {337}},\ \bibinfo {pages}
  {453} (\bibinfo {year} {2012})}\BibitemShut {NoStop}%
\bibitem [{\citenamefont {Agarwal}\ and\ \citenamefont
  {Escobedo}(2011)}]{Agarwal2011}%
  \BibitemOpen
  \bibfield  {author} {\bibinfo {author} {\bibfnamefont {U.}~\bibnamefont
  {Agarwal}}\ and\ \bibinfo {author} {\bibfnamefont {F.~A.}\ \bibnamefont
  {Escobedo}},\ }\href {\doibase 10.1038/nmat2959} {\bibfield  {journal}
  {\bibinfo  {journal} {Nat. Mater.}\ }\textbf {\bibinfo {volume} {10}},\
  \bibinfo {pages} {230} (\bibinfo {year} {2011})}\BibitemShut {NoStop}%
\bibitem [{\citenamefont {Teich}\ \emph {et~al.}(2016)\citenamefont {Teich},
  \citenamefont {van Anders}, \citenamefont {Klotsa}, \citenamefont
  {Dshemuchadse},\ and\ \citenamefont {Glotzer}}]{Teich2016}%
  \BibitemOpen
  \bibfield  {author} {\bibinfo {author} {\bibfnamefont {E.~G.}\ \bibnamefont
  {Teich}}, \bibinfo {author} {\bibfnamefont {G.}~\bibnamefont {van Anders}},
  \bibinfo {author} {\bibfnamefont {D.}~\bibnamefont {Klotsa}}, \bibinfo
  {author} {\bibfnamefont {J.}~\bibnamefont {Dshemuchadse}}, \ and\ \bibinfo
  {author} {\bibfnamefont {S.~C.}\ \bibnamefont {Glotzer}},\ }\href {\doibase
  10.1073/pnas.1524875113} {\bibfield  {journal} {\bibinfo  {journal} {Proc.
  Natl. Acad. Sci.}\ }\textbf {\bibinfo {volume} {113}},\ \bibinfo {pages}
  {E669} (\bibinfo {year} {2016})}\BibitemShut {NoStop}%
\bibitem [{\citenamefont {Lu}\ and\ \citenamefont {Weitz}(2013)}]{Lu2013}%
  \BibitemOpen
  \bibfield  {author} {\bibinfo {author} {\bibfnamefont {P.~J.}\ \bibnamefont
  {Lu}}\ and\ \bibinfo {author} {\bibfnamefont {D.~A.}\ \bibnamefont {Weitz}},\
  }\href {\doibase 10.1146/annurev-conmatphys-030212-184213} {\bibfield
  {journal} {\bibinfo  {journal} {Annual Review of Condensed Matter Physics}\
  }\textbf {\bibinfo {volume} {4}},\ \bibinfo {pages} {217} (\bibinfo {year}
  {2013})}\BibitemShut {NoStop}%
\bibitem [{\citenamefont {Nagel}(2017)}]{Nagel2017}%
  \BibitemOpen
  \bibfield  {author} {\bibinfo {author} {\bibfnamefont {S.~R.}\ \bibnamefont
  {Nagel}},\ }\href {\doibase 10.1103/RevModPhys.89.025002} {\bibfield
  {journal} {\bibinfo  {journal} {Rev. Mod. Phys.}\ }\textbf {\bibinfo {volume}
  {89}},\ \bibinfo {pages} {025002} (\bibinfo {year} {2017})}\BibitemShut
  {NoStop}%
\bibitem [{\citenamefont {John}\ and\ \citenamefont
  {Escobedo}(2005)}]{John2005}%
  \BibitemOpen
  \bibfield  {author} {\bibinfo {author} {\bibfnamefont {B.~S.}\ \bibnamefont
  {John}}\ and\ \bibinfo {author} {\bibfnamefont {F.~A.}\ \bibnamefont
  {Escobedo}},\ }\href@noop {} {\bibfield  {journal} {\bibinfo  {journal} {J.
  Phys. Chem. B}\ }\textbf {\bibinfo {volume} {109}},\ \bibinfo {pages} {23008}
  (\bibinfo {year} {2005})}\BibitemShut {NoStop}%
\bibitem [{\citenamefont {Khadilkar}\ and\ \citenamefont
  {Escobedo}(2012)}]{Khadilkar2012}%
  \BibitemOpen
  \bibfield  {author} {\bibinfo {author} {\bibfnamefont {M.~R.}\ \bibnamefont
  {Khadilkar}}\ and\ \bibinfo {author} {\bibfnamefont {F.~A.}\ \bibnamefont
  {Escobedo}},\ }\href@noop {} {\bibfield  {journal} {\bibinfo  {journal} {J.
  Chem. Phys.}\ }\textbf {\bibinfo {volume} {137}},\ \bibinfo {pages} {194907}
  (\bibinfo {year} {2012})}\BibitemShut {NoStop}%
\bibitem [{\citenamefont {Metropolis}\ \emph {et~al.}(1953)\citenamefont
  {Metropolis}, \citenamefont {Rosenbluth}, \citenamefont {Rosenbluth},
  \citenamefont {Teller},\ and\ \citenamefont {Teller}}]{Metropolis1953}%
  \BibitemOpen
  \bibfield  {author} {\bibinfo {author} {\bibfnamefont {N.}~\bibnamefont
  {Metropolis}}, \bibinfo {author} {\bibfnamefont {A.~W.}\ \bibnamefont
  {Rosenbluth}}, \bibinfo {author} {\bibfnamefont {M.~N.}\ \bibnamefont
  {Rosenbluth}}, \bibinfo {author} {\bibfnamefont {A.~H.}\ \bibnamefont
  {Teller}}, \ and\ \bibinfo {author} {\bibfnamefont {E.}~\bibnamefont
  {Teller}},\ }\href {\doibase 10.1063/1.1699114} {\bibfield  {journal}
  {\bibinfo  {journal} {J. Chem. Phys.}\ }\textbf {\bibinfo {volume} {21}},\
  \bibinfo {pages} {1087} (\bibinfo {year} {1953})}\BibitemShut {NoStop}%
\bibitem [{\citenamefont {Wood}(1968)}]{Wood1968}%
  \BibitemOpen
  \bibfield  {author} {\bibinfo {author} {\bibfnamefont {W.}~\bibnamefont
  {Wood}},\ }\href {\doibase 10.1063/1.1667938} {\bibfield  {journal} {\bibinfo
   {journal} {J. Chem. Phys.}\ }\textbf {\bibinfo {volume} {48}},\ \bibinfo
  {pages} {415} (\bibinfo {year} {1968})}\BibitemShut {NoStop}%
\bibitem [{\citenamefont {Wood}(1970)}]{Wood1970}%
  \BibitemOpen
  \bibfield  {author} {\bibinfo {author} {\bibfnamefont {W.}~\bibnamefont
  {Wood}},\ }\href {\doibase 10.1063/1.1673047} {\bibfield  {journal} {\bibinfo
   {journal} {J. Chem. Phys.}\ }\textbf {\bibinfo {volume} {52}},\ \bibinfo
  {pages} {729} (\bibinfo {year} {1970})}\BibitemShut {NoStop}%
\bibitem [{\citenamefont {Parrinello}\ and\ \citenamefont
  {Rahman}(1980)}]{Parrinello1980}%
  \BibitemOpen
  \bibfield  {author} {\bibinfo {author} {\bibfnamefont {M.}~\bibnamefont
  {Parrinello}}\ and\ \bibinfo {author} {\bibfnamefont {A.}~\bibnamefont
  {Rahman}},\ }\href {\doibase 10.1103/PhysRevLett.45.1196} {\bibfield
  {journal} {\bibinfo  {journal} {Phys. Rev. Lett.}\ }\textbf {\bibinfo
  {volume} {45}},\ \bibinfo {pages} {1196} (\bibinfo {year}
  {1980})}\BibitemShut {NoStop}%
\bibitem [{\citenamefont {Parrinello}\ and\ \citenamefont
  {Rahman}(1981)}]{Parrinello1981}%
  \BibitemOpen
  \bibfield  {author} {\bibinfo {author} {\bibfnamefont {M.}~\bibnamefont
  {Parrinello}}\ and\ \bibinfo {author} {\bibfnamefont {A.}~\bibnamefont
  {Rahman}},\ }\href {\doibase 10.1063/1.328693} {\bibfield  {journal}
  {\bibinfo  {journal} {J. Appl. Phys.}\ }\textbf {\bibinfo {volume} {52}},\
  \bibinfo {pages} {7182} (\bibinfo {year} {1981})}\BibitemShut {NoStop}%
\bibitem [{\citenamefont {Najafabadi}\ and\ \citenamefont
  {Yip}(1983)}]{Najafabadi1983}%
  \BibitemOpen
  \bibfield  {author} {\bibinfo {author} {\bibfnamefont {R.}~\bibnamefont
  {Najafabadi}}\ and\ \bibinfo {author} {\bibfnamefont {S.}~\bibnamefont
  {Yip}},\ }\href {\doibase 10.1016/0036-9748(83)90283-1} {\bibfield  {journal}
  {\bibinfo  {journal} {Scr. Metall.}\ }\textbf {\bibinfo {volume} {17}},\
  \bibinfo {pages} {1199} (\bibinfo {year} {1983})}\BibitemShut {NoStop}%
\bibitem [{\citenamefont {Yashonath}\ and\ \citenamefont
  {Rao}(1985)}]{Yashonath1985}%
  \BibitemOpen
  \bibfield  {author} {\bibinfo {author} {\bibfnamefont {S.}~\bibnamefont
  {Yashonath}}\ and\ \bibinfo {author} {\bibfnamefont {C.}~\bibnamefont
  {Rao}},\ }\href {\doibase 10.1080/00268978500100201} {\bibfield  {journal}
  {\bibinfo  {journal} {Mol. Phys.}\ }\textbf {\bibinfo {volume} {54}},\
  \bibinfo {pages} {245} (\bibinfo {year} {1985})}\BibitemShut {NoStop}%
\bibitem [{\citenamefont {Filion}\ \emph {et~al.}(2009)\citenamefont {Filion},
  \citenamefont {Marechal}, \citenamefont {van Oorschot}, \citenamefont {Pelt},
  \citenamefont {Smallenburg},\ and\ \citenamefont {Dijkstra}}]{Filion2009}%
  \BibitemOpen
  \bibfield  {author} {\bibinfo {author} {\bibfnamefont {L.}~\bibnamefont
  {Filion}}, \bibinfo {author} {\bibfnamefont {M.}~\bibnamefont {Marechal}},
  \bibinfo {author} {\bibfnamefont {B.}~\bibnamefont {van Oorschot}}, \bibinfo
  {author} {\bibfnamefont {D.}~\bibnamefont {Pelt}}, \bibinfo {author}
  {\bibfnamefont {F.}~\bibnamefont {Smallenburg}}, \ and\ \bibinfo {author}
  {\bibfnamefont {M.}~\bibnamefont {Dijkstra}},\ }\href {\doibase
  10.1103/PhysRevLett.103.188302} {\bibfield  {journal} {\bibinfo  {journal}
  {Phys. Rev. Lett.}\ }\textbf {\bibinfo {volume} {103}},\ \bibinfo {pages}
  {188302} (\bibinfo {year} {2009})}\BibitemShut {NoStop}%
\bibitem [{\citenamefont {Plimpton}(1995)}]{Plimpton1995}%
  \BibitemOpen
  \bibfield  {author} {\bibinfo {author} {\bibfnamefont {S.}~\bibnamefont
  {Plimpton}},\ }\href {\doibase 10.1006/jcph.1995.1039} {\bibfield  {journal}
  {\bibinfo  {journal} {Journal of Computational Physics}\ }\textbf {\bibinfo
  {volume} {117}},\ \bibinfo {pages} {1} (\bibinfo {year} {1995})}\BibitemShut
  {NoStop}%
\bibitem [{\citenamefont {Martyna}\ \emph {et~al.}(1992)\citenamefont
  {Martyna}, \citenamefont {Klein},\ and\ \citenamefont
  {Tuckerman}}]{Martyna1992}%
  \BibitemOpen
  \bibfield  {author} {\bibinfo {author} {\bibfnamefont {G.~J.}\ \bibnamefont
  {Martyna}}, \bibinfo {author} {\bibfnamefont {M.~L.}\ \bibnamefont {Klein}},
  \ and\ \bibinfo {author} {\bibfnamefont {M.}~\bibnamefont {Tuckerman}},\
  }\href {\doibase 10.1063/1.463940} {\bibfield  {journal} {\bibinfo  {journal}
  {The Journal of Chemical Physics}\ }\textbf {\bibinfo {volume} {97}},\
  \bibinfo {pages} {2635} (\bibinfo {year} {1992})}\BibitemShut {NoStop}%
\bibitem [{\citenamefont {Kamberaj}\ \emph {et~al.}(2005)\citenamefont
  {Kamberaj}, \citenamefont {Low},\ and\ \citenamefont {Neal}}]{Kamberaj2005}%
  \BibitemOpen
  \bibfield  {author} {\bibinfo {author} {\bibfnamefont {H.}~\bibnamefont
  {Kamberaj}}, \bibinfo {author} {\bibfnamefont {R.~J.}\ \bibnamefont {Low}}, \
  and\ \bibinfo {author} {\bibfnamefont {M.~P.}\ \bibnamefont {Neal}},\ }\href
  {\doibase 10.1063/1.1906216} {\bibfield  {journal} {\bibinfo  {journal} {The
  Journal of Chemical Physics}\ }\textbf {\bibinfo {volume} {122}},\ \bibinfo
  {pages} {224114} (\bibinfo {year} {2005})}\BibitemShut {NoStop}%
\bibitem [{\citenamefont {Vesely}(1982)}]{Vesely1982}%
  \BibitemOpen
  \bibfield  {author} {\bibinfo {author} {\bibfnamefont {F.~J.}\ \bibnamefont
  {Vesely}},\ }\href {\doibase 10.1016/0021-9991(82)90080-8} {\bibfield
  {journal} {\bibinfo  {journal} {J. Comput. Phys.}\ }\textbf {\bibinfo
  {volume} {47}},\ \bibinfo {pages} {291} (\bibinfo {year} {1982})}\BibitemShut
  {NoStop}%
\bibitem [{\citenamefont {Wang}\ \emph {et~al.}(2005)\citenamefont {Wang},
  \citenamefont {Teitel},\ and\ \citenamefont {Dellago}}]{Wang2005}%
  \BibitemOpen
  \bibfield  {author} {\bibinfo {author} {\bibfnamefont {Y.}~\bibnamefont
  {Wang}}, \bibinfo {author} {\bibfnamefont {S.}~\bibnamefont {Teitel}}, \ and\
  \bibinfo {author} {\bibfnamefont {C.}~\bibnamefont {Dellago}},\ }\href
  {\doibase 10.1063/1.1917756} {\bibfield  {journal} {\bibinfo  {journal} {J.
  Chem. Phys.}\ }\textbf {\bibinfo {volume} {122}},\ \bibinfo {pages} {214722}
  (\bibinfo {year} {2005})}\BibitemShut {NoStop}%
\bibitem [{\citenamefont {Chen}\ \emph {et~al.}(2014)\citenamefont {Chen},
  \citenamefont {Jiao},\ and\ \citenamefont {Torquato}}]{Chen2014}%
  \BibitemOpen
  \bibfield  {author} {\bibinfo {author} {\bibfnamefont {D.}~\bibnamefont
  {Chen}}, \bibinfo {author} {\bibfnamefont {Y.}~\bibnamefont {Jiao}}, \ and\
  \bibinfo {author} {\bibfnamefont {S.}~\bibnamefont {Torquato}},\ }\href
  {\doibase 10.1021/jp5010133} {\bibfield  {journal} {\bibinfo  {journal} {J.
  Phys. Chem. B}\ }\textbf {\bibinfo {volume} {118}},\ \bibinfo {pages} {7981}
  (\bibinfo {year} {2014})}\BibitemShut {NoStop}%
\bibitem [{\citenamefont {Boles}\ \emph {et~al.}(2016)\citenamefont {Boles},
  \citenamefont {Engel},\ and\ \citenamefont {Talapin}}]{Boles2016}%
  \BibitemOpen
  \bibfield  {author} {\bibinfo {author} {\bibfnamefont {M.~A.}\ \bibnamefont
  {Boles}}, \bibinfo {author} {\bibfnamefont {M.}~\bibnamefont {Engel}}, \ and\
  \bibinfo {author} {\bibfnamefont {D.~V.}\ \bibnamefont {Talapin}},\ }\href
  {\doibase 10.1021/acs.chemrev.6b00196} {\bibfield  {journal} {\bibinfo
  {journal} {Chem. Rev.}\ }\textbf {\bibinfo {volume} {116}},\ \bibinfo {pages}
  {11220} (\bibinfo {year} {2016})}\BibitemShut {NoStop}%
\bibitem [{\citenamefont {Zhang}\ \emph {et~al.}(2013)\citenamefont {Zhang},
  \citenamefont {Lu}, \citenamefont {Yager}, \citenamefont {Van Der~Lelie},\
  and\ \citenamefont {Gang}}]{Zhang2013}%
  \BibitemOpen
  \bibfield  {author} {\bibinfo {author} {\bibfnamefont {Y.}~\bibnamefont
  {Zhang}}, \bibinfo {author} {\bibfnamefont {F.}~\bibnamefont {Lu}}, \bibinfo
  {author} {\bibfnamefont {K.~G.}\ \bibnamefont {Yager}}, \bibinfo {author}
  {\bibfnamefont {D.}~\bibnamefont {Van Der~Lelie}}, \ and\ \bibinfo {author}
  {\bibfnamefont {O.}~\bibnamefont {Gang}},\ }\href {\doibase
  10.1038/nnano.2013.209} {\bibfield  {journal} {\bibinfo  {journal} {Nat.
  Nanotechnol.}\ }\textbf {\bibinfo {volume} {8}},\ \bibinfo {pages} {865}
  (\bibinfo {year} {2013})}\BibitemShut {NoStop}%
\bibitem [{\citenamefont {Ye}\ \emph {et~al.}(2016)\citenamefont {Ye},
  \citenamefont {Pearson}, \citenamefont {Cordeau}, \citenamefont {Mefford},\
  and\ \citenamefont {Crawford}}]{Ye2016}%
  \BibitemOpen
  \bibfield  {author} {\bibinfo {author} {\bibfnamefont {L.}~\bibnamefont
  {Ye}}, \bibinfo {author} {\bibfnamefont {T.}~\bibnamefont {Pearson}},
  \bibinfo {author} {\bibfnamefont {Y.}~\bibnamefont {Cordeau}}, \bibinfo
  {author} {\bibfnamefont {O.}~\bibnamefont {Mefford}}, \ and\ \bibinfo
  {author} {\bibfnamefont {T.}~\bibnamefont {Crawford}},\ }\href {\doibase
  10.1038/srep23145} {\bibfield  {journal} {\bibinfo  {journal} {Sci. Rep.}\
  }\textbf {\bibinfo {volume} {6}},\ \bibinfo {pages} {23145} (\bibinfo {year}
  {2016})}\BibitemShut {NoStop}%
\bibitem [{\citenamefont {Frenkel}\ and\ \citenamefont
  {Smit}(2001)}]{Frenkel2001}%
  \BibitemOpen
  \bibfield  {author} {\bibinfo {author} {\bibfnamefont {D.}~\bibnamefont
  {Frenkel}}\ and\ \bibinfo {author} {\bibfnamefont {B.}~\bibnamefont {Smit}},\
  }\href
  {https://www.elsevier.com/books/understanding-molecular-simulation/frenkel/978-0-12-267351-1}
  {\emph {\bibinfo {title} {Understanding molecular simulation: from algorithms
  to applications}}}\ (\bibinfo  {publisher} {Academic Press, San Diego},\
  \bibinfo {year} {2001})\BibitemShut {NoStop}%
\end{thebibliography}%

\appendix

\section{Monte Carlo moves to break pairs of icosahedra}

At low temperatures the particles can form face-to-face contacts, which are locally favorable in energy, but globally incompatible with the most stable crystal structure. To equilibrate such a system we need to efficiently sample all the possible configurations and it is therefore necessary to be able to break up these contacts.
We do this by using trial moves where we randomly choose two icosahedra and displace them out of the attraction range of each other. In the following we explain how the detailed balance is preserved for these moves.

We use the basic Metropolis scheme with the acceptance rule~\cite{Frenkel2001}
\begin{equation}
 {{\rm acc}({\rm o}\to{\rm n})\over{\rm acc}({\rm n}\to{\rm o})}=\min\left[1,f{\rm e}^{-\beta\left\{U({\rm n})-U({\rm o})\right\}}\right]
\end{equation}
which takes into account different probabilities of generating configurations through a factor $f\equiv\alpha({\rm n}\to {\rm o})/\alpha({\rm o}\to {\rm n})$.
We define a sphere of influence by a distance $d_{\rm infl}=7.0\,\sigma$. We denote the volume of the corresponding sphere by $V_{\rm s}={4\over3}\pi d_{\rm infl}^3$ and the volume of the simulation box by $V_{\rm box}$.

\begin{figure}
  \centering
  \includegraphics{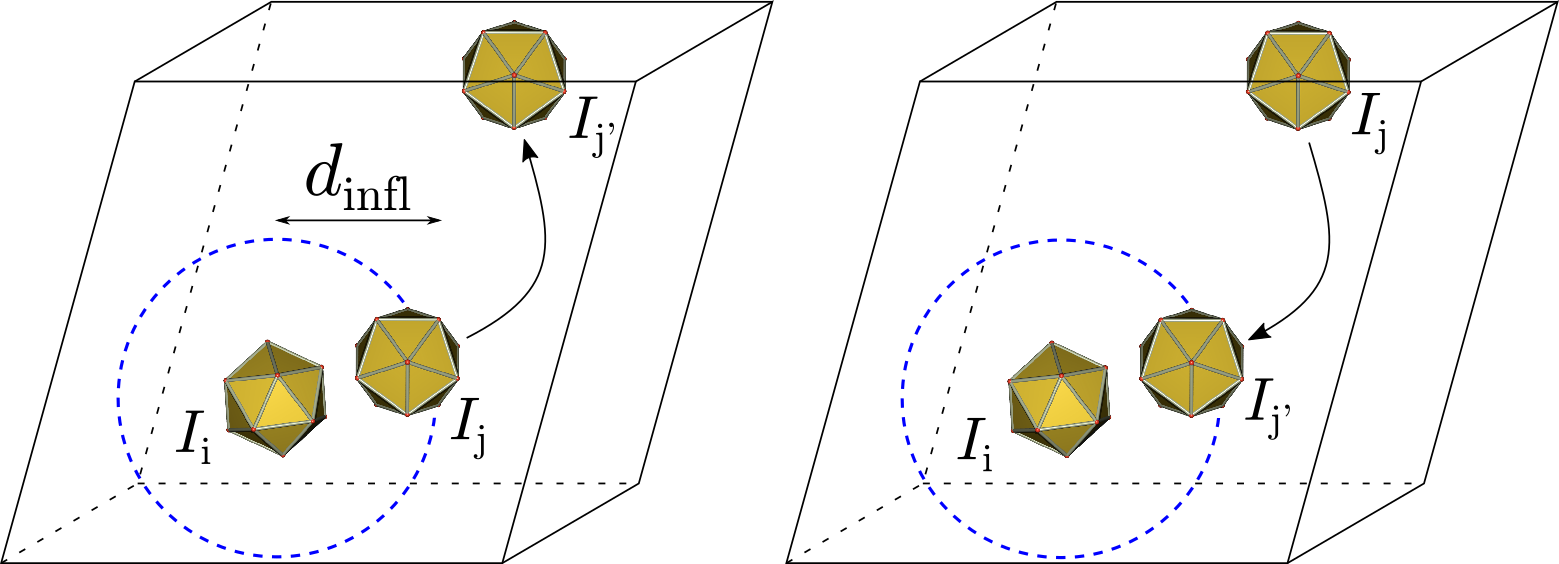}
  \caption{In the Monte Carlo move to break pairs of icosahedra we consider two different scenarios to preserve detailed balance.}
  \label{fig:move}
\end{figure}

We then randomly select two icosahedra, $I_i$ and $I_j$, as shown in Fig.~\ref{fig:move}, and check whether $I_j$ lies inside the sphere of influence of $I_i$. In this case we move $I_j$ out of sphere with volume $V_{\rm s}$ around $I_i$ with the corresponding acceptance factor $f=\left({V_{\rm box}-V_{\rm s}\over V_{\rm box}}\right)/\left({V_{\rm s}\over V_{\rm box}}\right)=V_{\rm box}/ V_{\rm s}-1$. In case when $I_j$ lies outside the sphere of influence of $I_i$ we move $I_j$ randomly inside the sphere of influence of $I_i$ and apply a reverse acceptance factor $f=V_{\rm s}/\left(V_{\rm box}-V_{\rm s}\right)$.

\section{Analysis}

\begin{figure*}
  \centering
  \subfloat[]{{\includegraphics[width=.66\columnwidth]{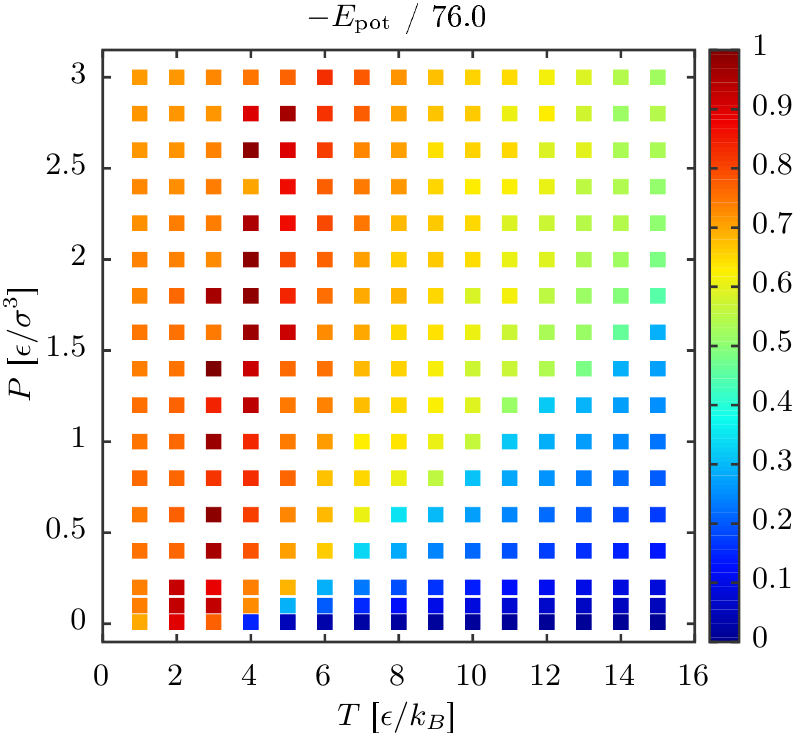}}\label{fig:analysis:a}}
  \subfloat[]{{\includegraphics[width=.66\columnwidth]{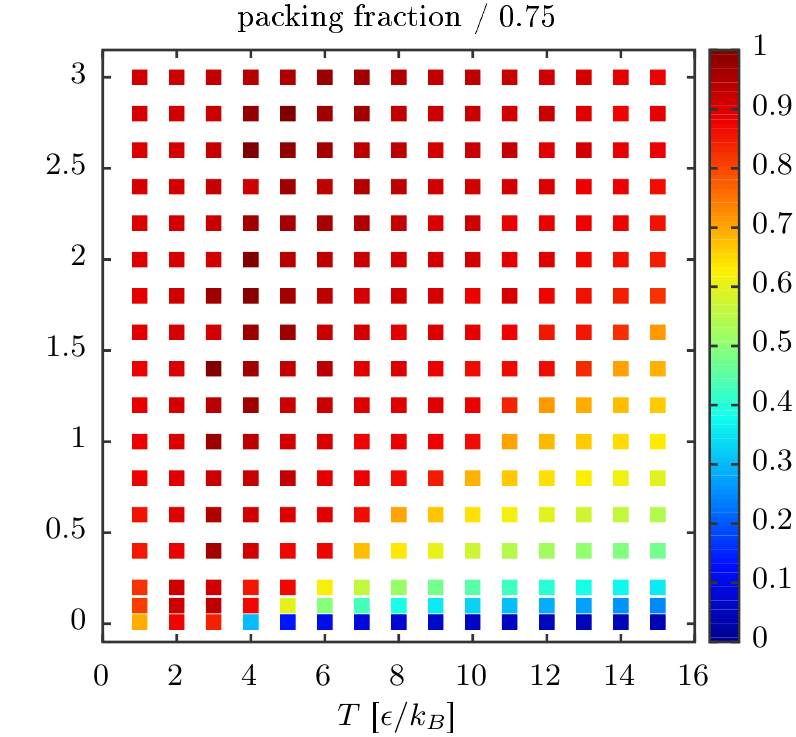}}\label{fig:analysis:b}}
  \subfloat[]{{\includegraphics[width=.66\columnwidth]{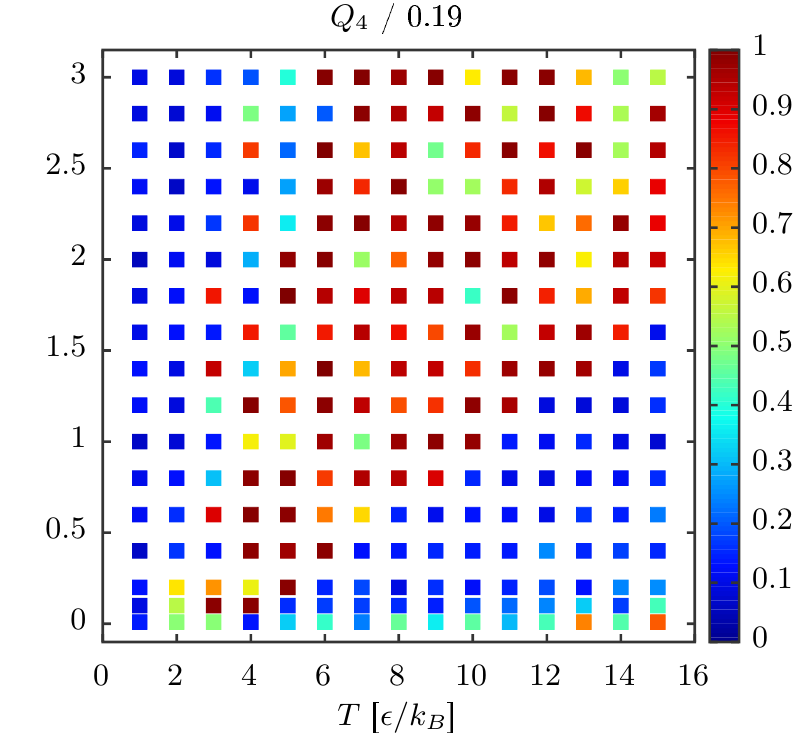}}\label{fig:analysis:c}}\\
  \subfloat[]{{\includegraphics[width=.66\columnwidth]{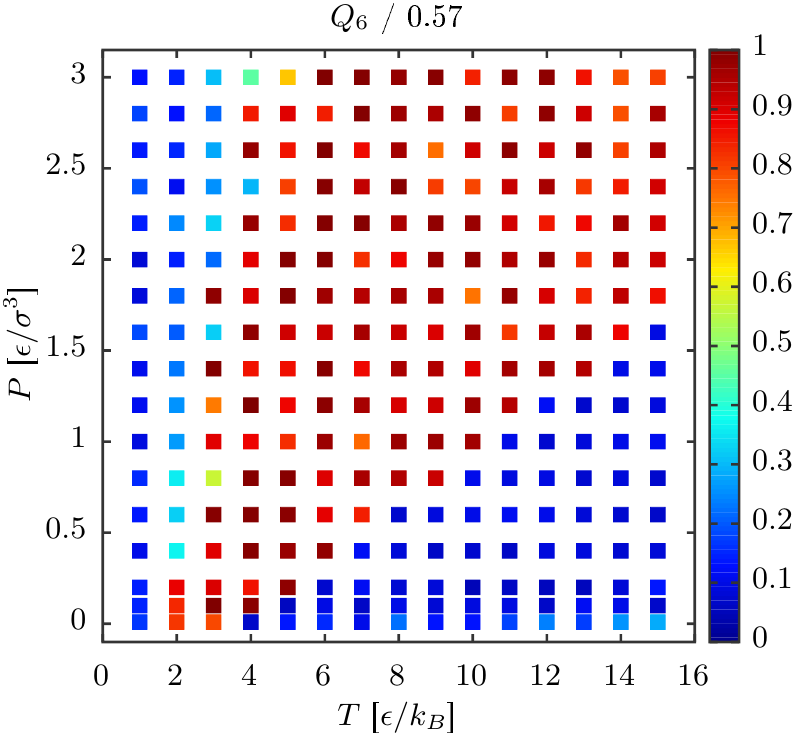}}\label{fig:analysis:d}}
  \subfloat[]{{\includegraphics[width=.66\columnwidth]{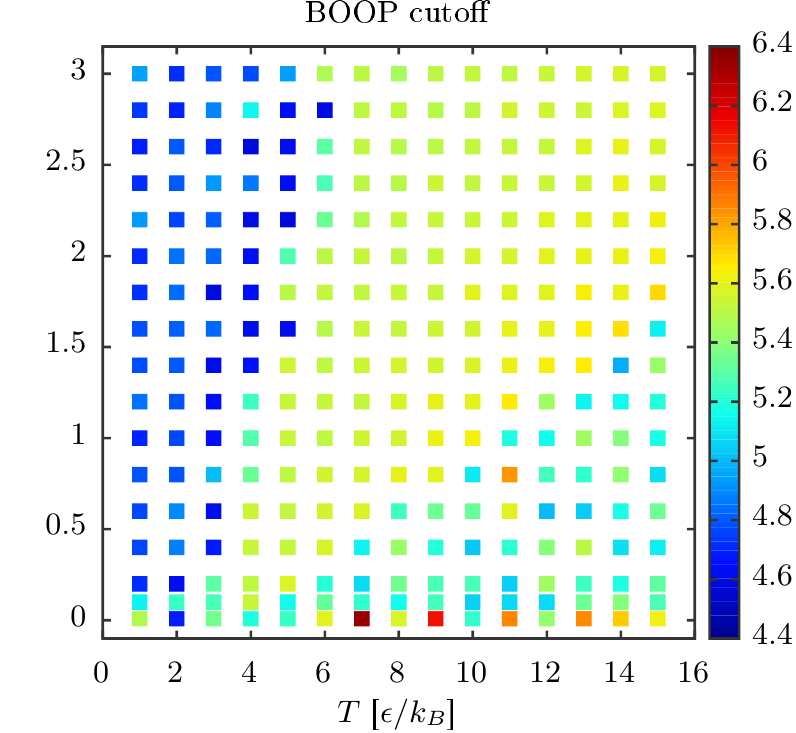}}\label{fig:analysis:e}}
  \subfloat[]{{\includegraphics[width=.66\columnwidth]{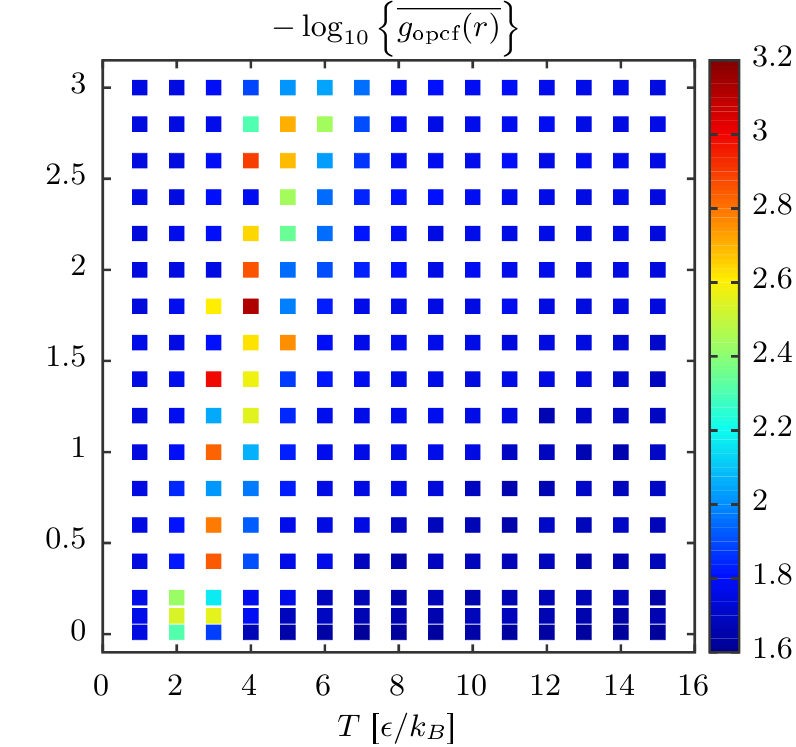}}\label{fig:analysis:f}}
  \caption{Properties and structural analysis of systems of 500 Mackay icosahedra after simulation at constant values of temperature and pressure starting from initially disordered fluid. Values of potential energy (a), packing fraction (b), bond-orientational order parameters $Q_4$ (c) and $Q_6$ (d), the position between the first two peaks in RDF used for BOOP calculations (e) and average value of OPCF (f) are averaged over 50 snapshots of the system.}
  \label{fig:analysis}
\end{figure*}

Several quantities that characterize the system are shown in Fig.~\ref{fig:analysis}.
The disordered regions are most clearly identified by their low values of bond-orientational order parameters in Figs.~\ref{fig:analysis:c} and \ref{fig:analysis:d}, which we used to distinguish between positionally ordered and disordered phases according to the chosen threshold values of 0.14 and 0.42 for $Q_4$ and $Q_6$ respectively.
Orientational order was determined according to the threshold value of 0.007 for the average value of the OPCF calculated with normals of the faces of icosahedra as characteristic vectors.
Additionally, the comparison of the peaks in radial distribution function with that of ideal FCC, BCC and DLP lattices was used to more precisely categorize the positional order.

The region where crystal configurations form is clearly seen from their low potential energy values, their high packing fraction and low average values of orientational pair correlation function in Figs.~\ref{fig:analysis:a}, \ref{fig:analysis:b} and \ref{fig:analysis:f}.
The border between the regions with crystal and rotator crystal arrangements can be determined from the average value of orientational correlations in Fig.~\ref{fig:analysis:f}.
We also see that the disordered solid region that forms at low temperatures significantly differs from the liquid disordered region at high temperatures. At low temperatures configurations have notably larger packing fractions and lower potential energies than at higher temperatures where they form a fully disordered low density phase with a corresponding low potential energy.
From Fig.~\ref{fig:analysis:e} we furthermore see that the position between the first two peaks in the radial distribution function has clearly larger values in the high temperature disordered region than at low temperatures. This is because the peaks in RDF become less distinct and gradually disappear for the highly disordered, low density configurations that are stable at high temperatures.

\section{Results for different initial conditions}

\begin{figure*}
  \centering
  \subfloat[]{{\includegraphics[width=.66\columnwidth]{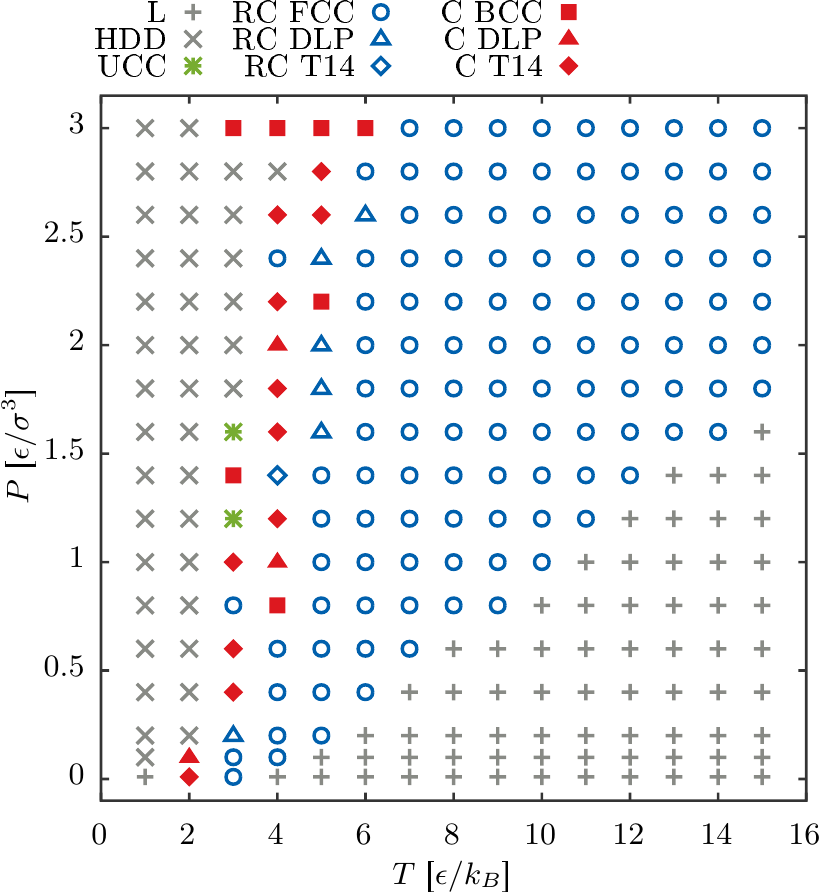}}\label{fig:diags1:a}}
  \subfloat[]{{\includegraphics[width=.66\columnwidth]{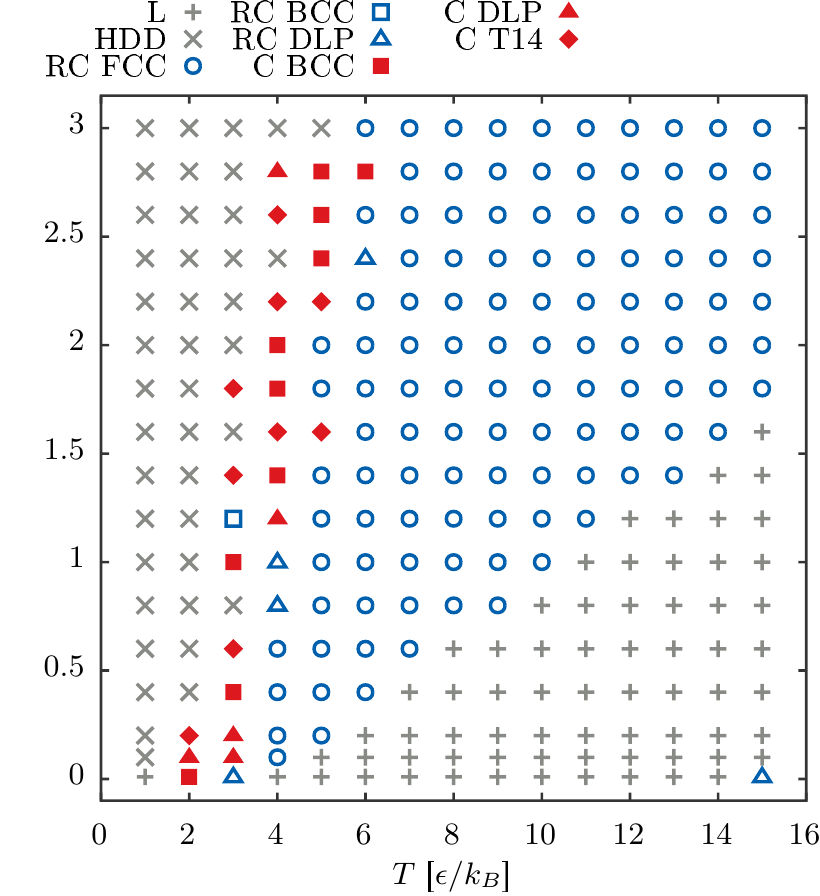}}\label{fig:diags1:b}}
  \subfloat[]{{\includegraphics[width=.66\columnwidth]{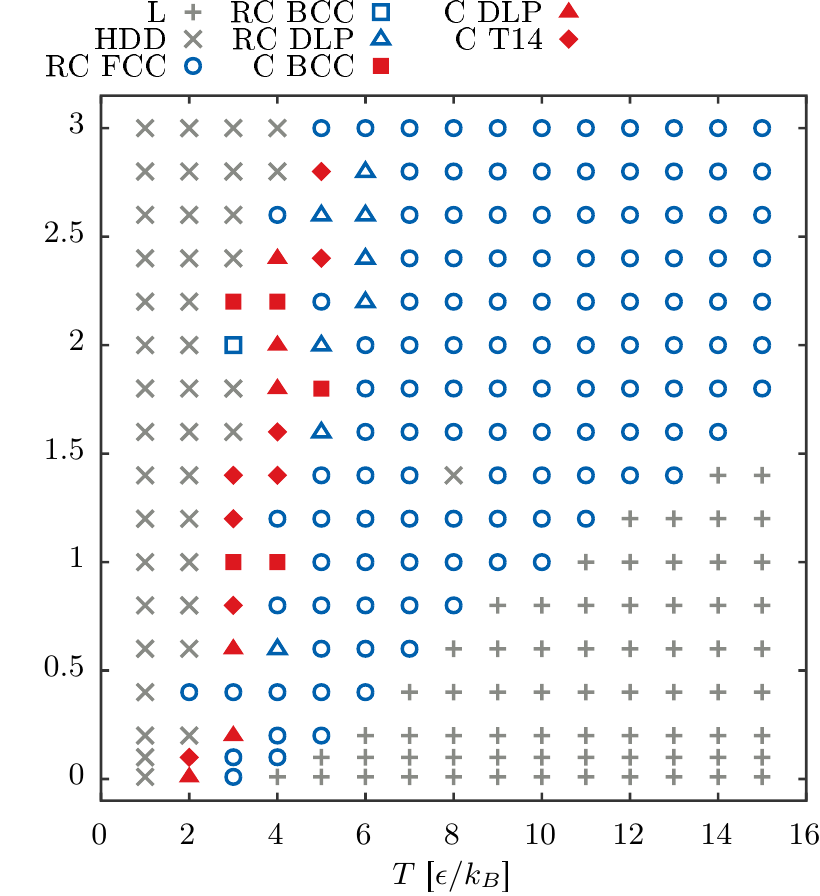}}\label{fig:diags1:c}}
  \caption{Scan of the phase diagram obtained by simulation of the systems of 500 Mackay icosahedra starting from three different sets of low density disordered initial configurations. We use the following symbols to denote different structures: {\color{col_l}\bm{$+$}} for a liquid, {\color{col_l}\bm{$\times$}} for a high-density disordered phase, {\color{col_lc}{\bf\plusovercross}} for a unequilibrated crystalline configuration, {\color{col_rc}\bm{$\circ$}} for a rotator crystal with FCC positional order, {\color{col_rc}\bm{$\square$}} for a rotator crystal with BCC positional order, {\color{col_rc}\bm{$\triangle$}} for a rotator crystal with DLP positional order, {\color{col_rc}\bm{$\Diamond$}} for a rotator crystal with T14 positional order, {\color{col_c}$\blacksquare$} for a crystal with BCC positional order, {\color{col_c}$\blacktriangle$} for a crystal with DLP positional order, {\color{col_c}$\blacklozenge$} for a crystal with T14 positional order. An analysis of the systems obtained in Fig.~\ref{fig:diags1:b} is shown in Fig.~\ref{fig:analysis}.}
  \label{fig:diags1}
\end{figure*}

\begin{figure*}
  \centering
  \subfloat[]{{\includegraphics[width=.66\columnwidth]{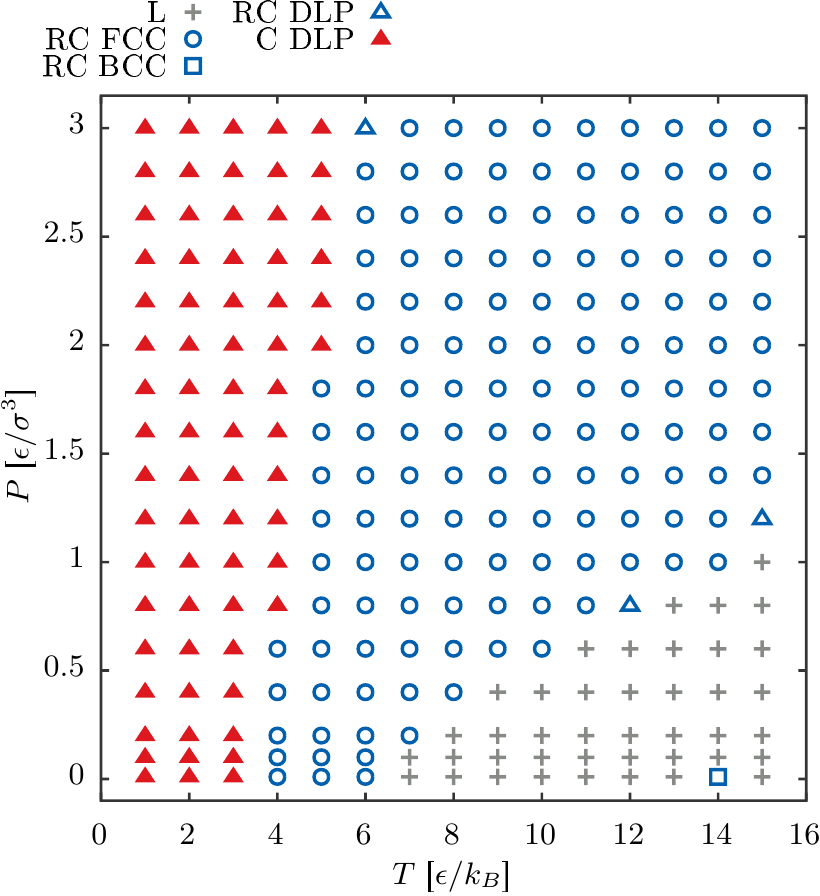}}\label{fig:diags2:a}}
  \subfloat[]{{\includegraphics[width=.66\columnwidth]{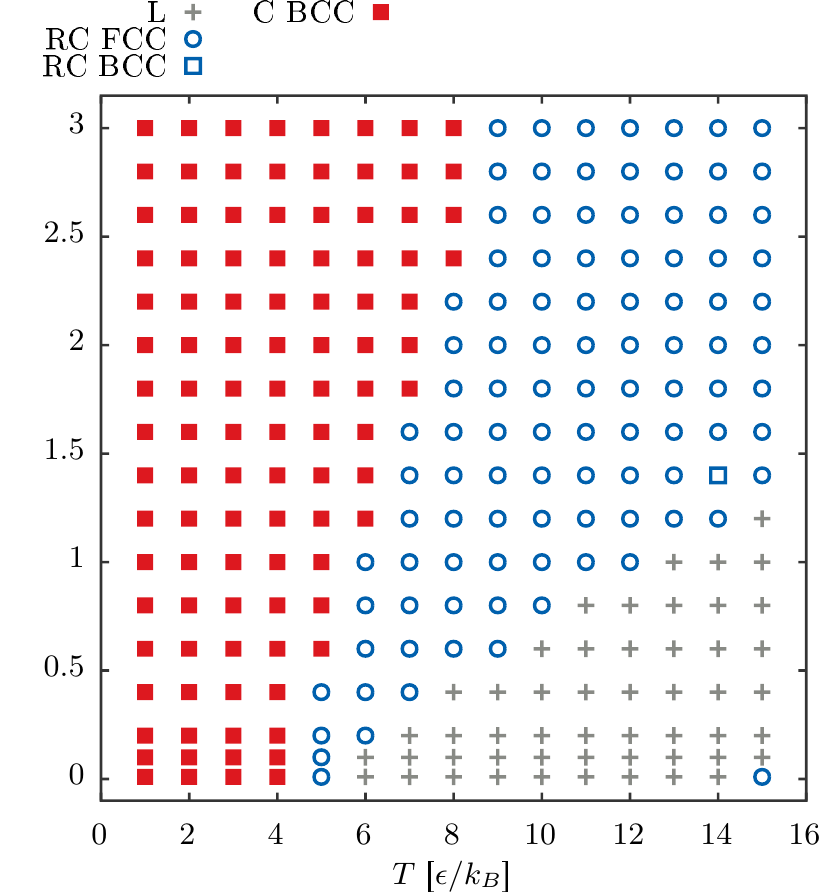}}\label{fig:diags2:b}}
  \subfloat[]{{\includegraphics[width=.66\columnwidth]{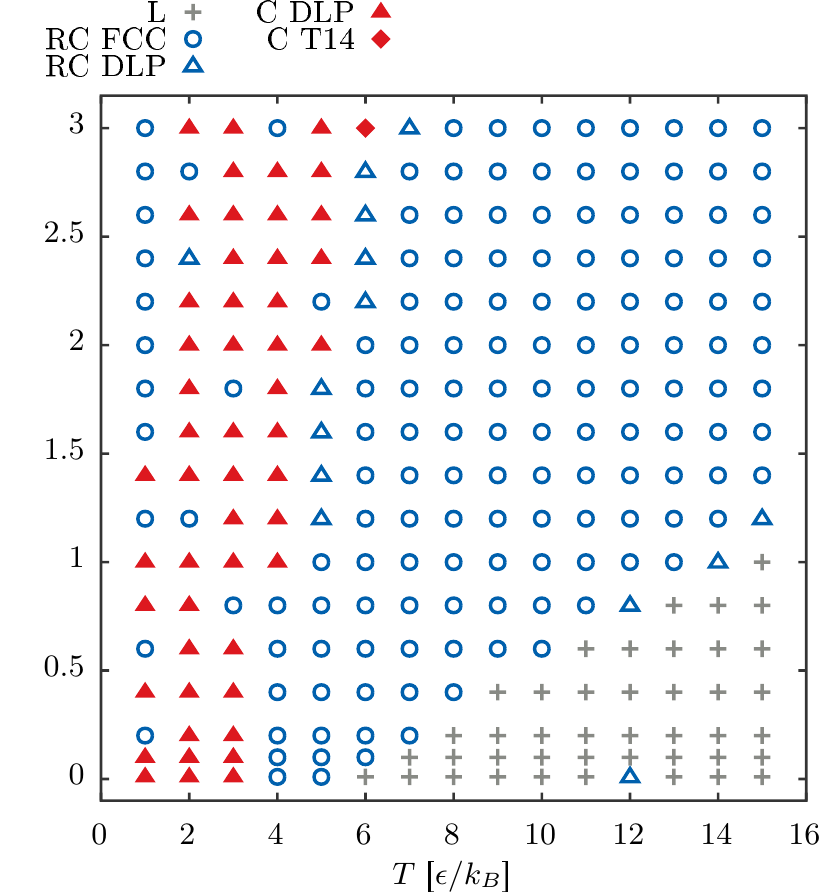}}\label{fig:diags2:c}}
  \caption{Scan of the phase diagram obtained by simulation of the systems of 500 Mackay icosahedra starting from three different initial configurations, DLP crystal (a), BCC crystal (b) and an FCC rotator crystal (c), all of which were obtained as a result in the diagrams shown in Fig.~\ref{fig:diags1}. We use the following symbols to denote different structures: {\color{col_l}\bm{$+$}} for a liquid, {\color{col_rc}\bm{$\circ$}} for a rotator crystal with FCC positional order, {\color{col_rc}\bm{$\square$}} for a rotator crystal with BCC positional order, {\color{col_rc}\bm{$\triangle$}} for a rotator crystal with DLP positional order, {\color{col_c}$\blacksquare$} for a crystal with BCC positional order, {\color{col_c}$\blacktriangle$} for a crystal with DLP positional order, {\color{col_c}$\blacklozenge$} for a crystal with T14 positional order.}
  \label{fig:diags2}
\end{figure*}

The results of the phase diagram scan are shown in Fig.~\ref{fig:diags1} for three different sets of disordered fluid initial configurations and in Fig.~\ref{fig:diags2} for simulations started from the BCC and DLP crystal lattices and an FCC rotator crystal phase.

\section{Equation of state}

\begin{figure*}
  \centering
  \subfloat[]{{\includegraphics[width=.66\columnwidth]{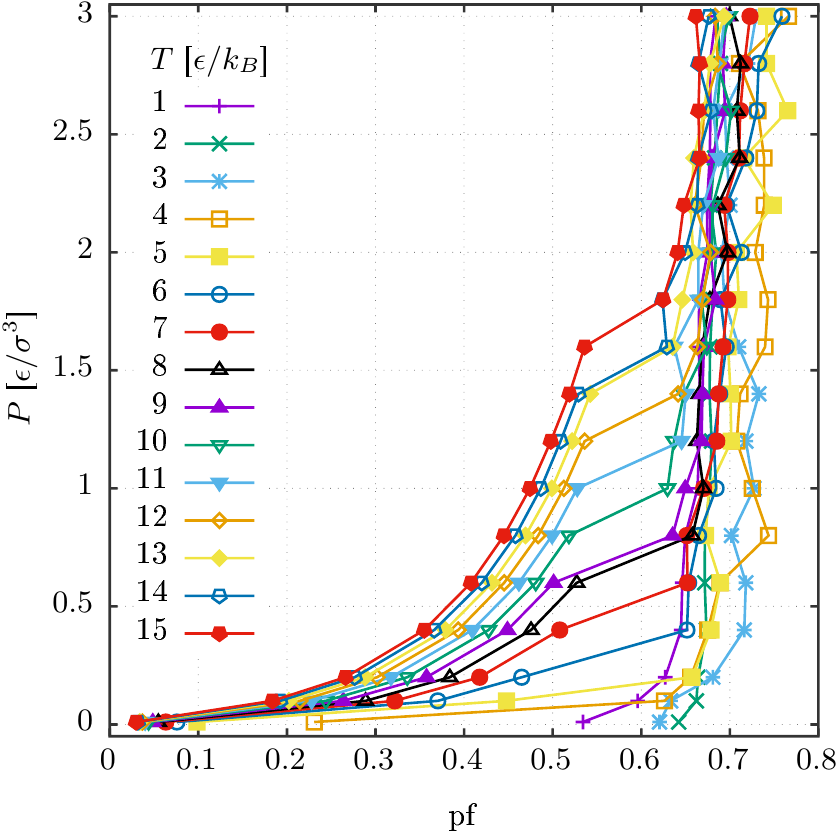}}\label{fig:eos1:a}}
  \subfloat[]{{\includegraphics[width=.66\columnwidth]{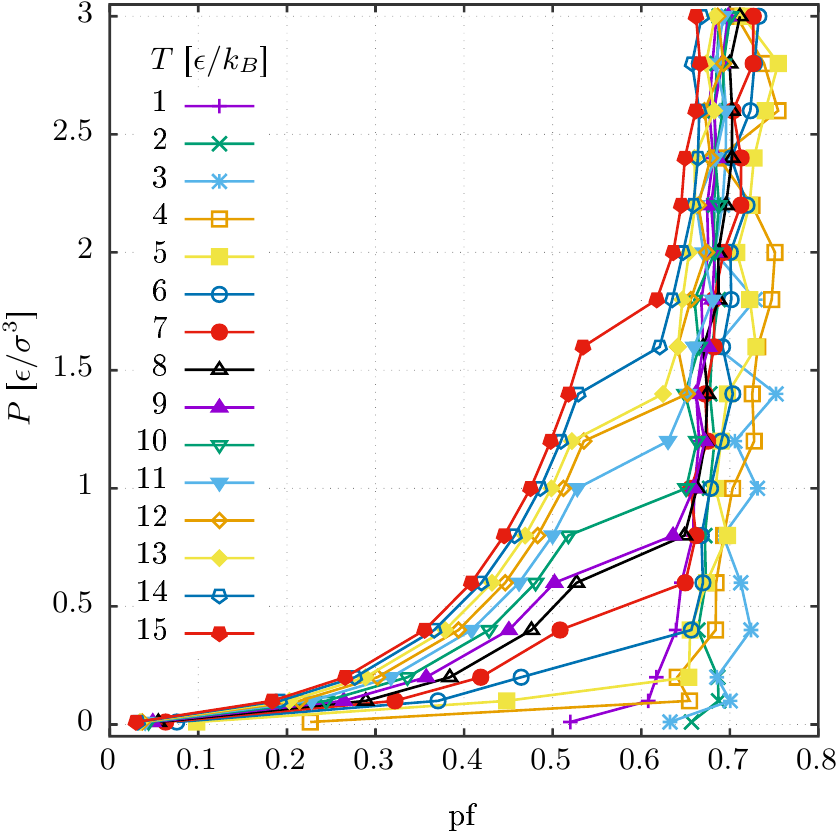}}\label{fig:eos1:b}}
  \subfloat[]{{\includegraphics[width=.66\columnwidth]{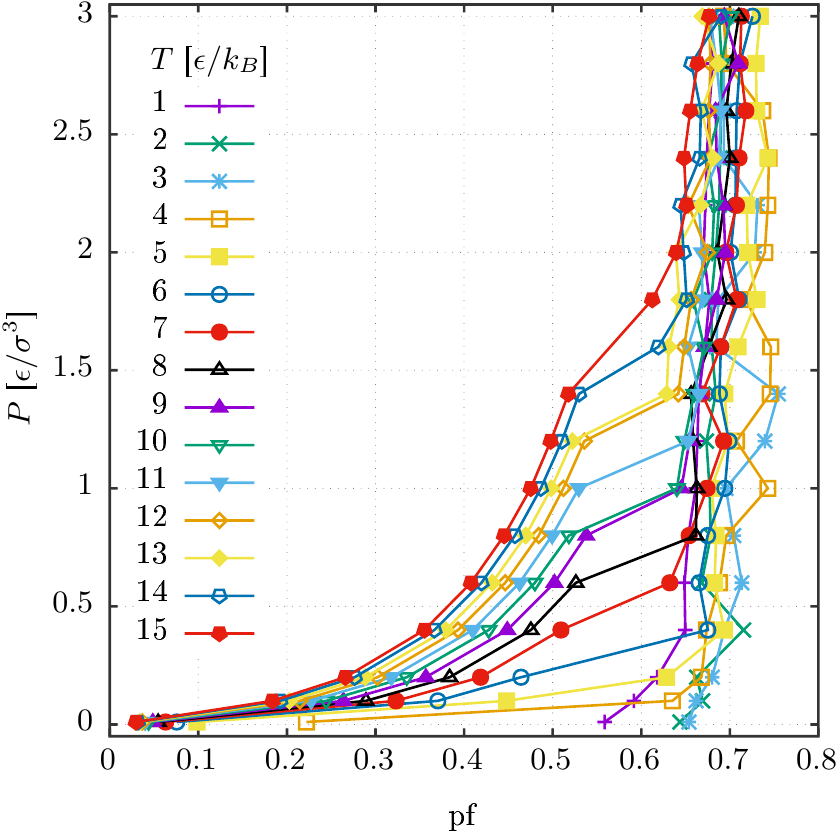}}\label{fig:eos1:c}}
  \caption{Equation of state curves in pressures vs. packing fraction representation at different temperatures corresponding to diagrams in Figs.~\ref{fig:diags1:a}, \ref{fig:diags1:b} and \ref{fig:diags1:c}.}
  \label{fig:eos1}
\end{figure*}

\begin{figure*}
  \centering
  \subfloat[]{{\includegraphics[width=.66\columnwidth]{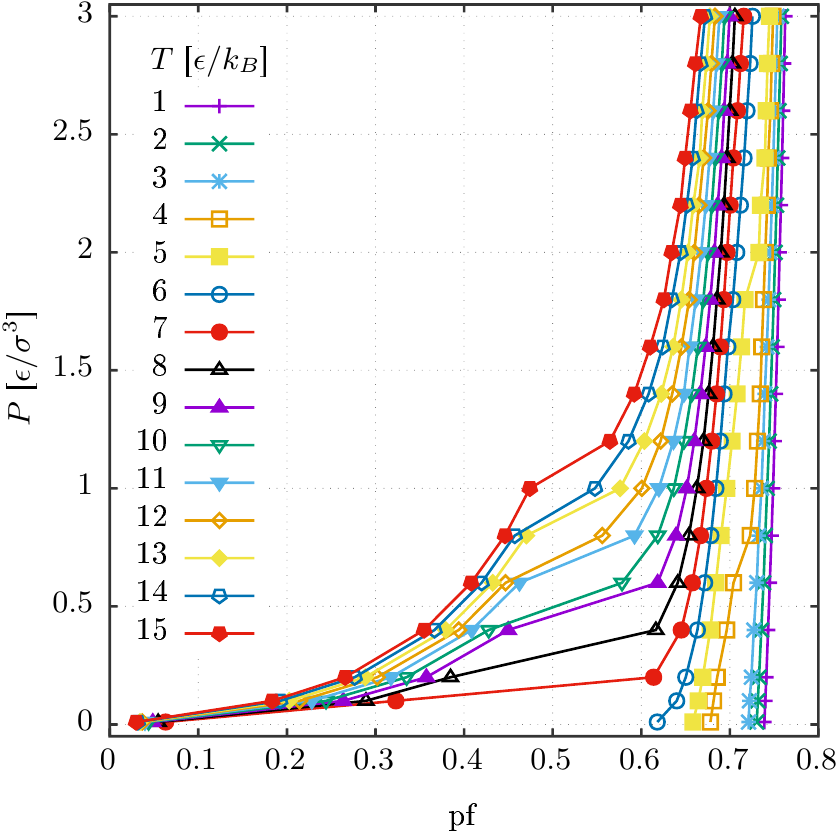}}\label{fig:eos:a}}
  \subfloat[]{{\includegraphics[width=.66\columnwidth]{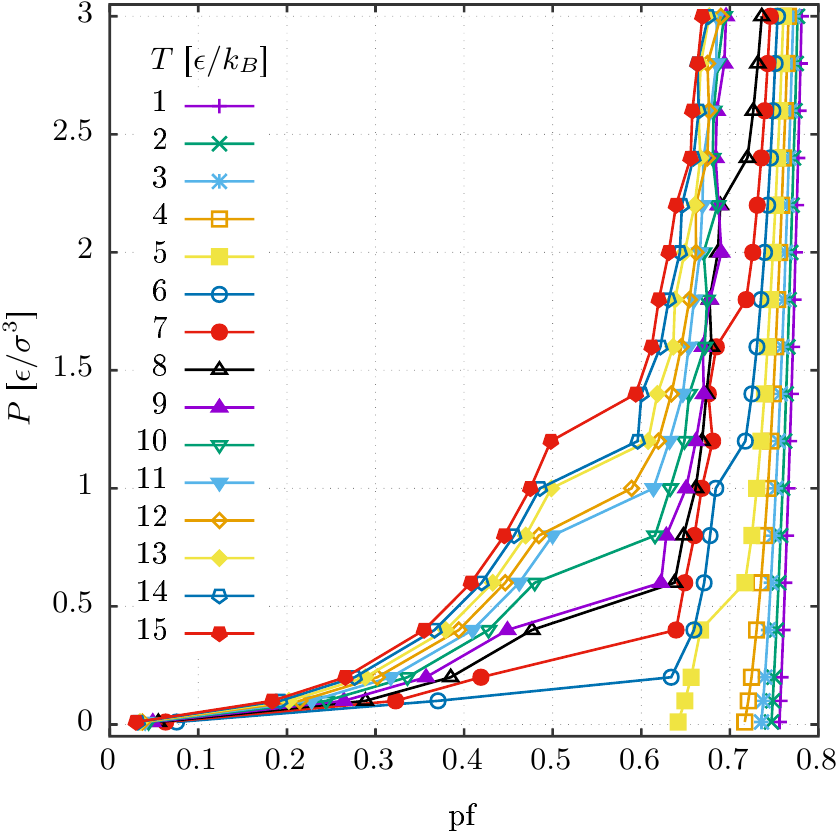}}\label{fig:eos:b}}
  \subfloat[]{{\includegraphics[width=.66\columnwidth]{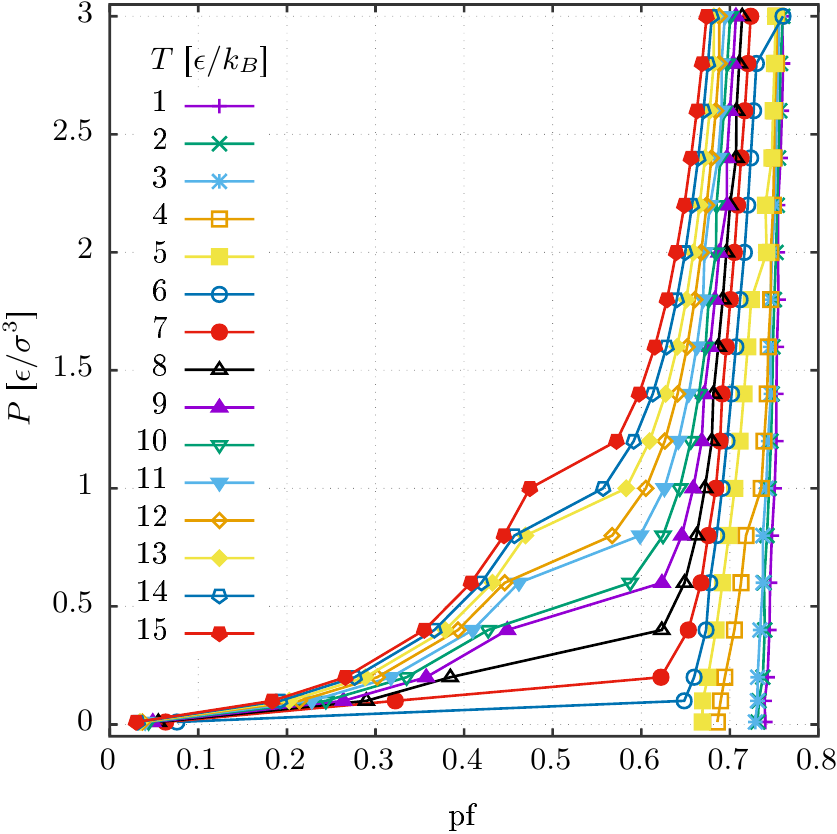}}\label{fig:eos:c}}
  \caption{Equation of state curves in pressures vs. packing fraction representation at different temperatures corresponding to diagrams in Figs.~\ref{fig:diags2:a}, \ref{fig:diags2:b} and \ref{fig:diags2:c}.}
  \label{fig:eos2}
\end{figure*}

The equations of state for the resulting phase diagrams are shown in Figs.~\ref{fig:eos1} and \ref{fig:eos2} as two-dimensional curves of pressures vs. packing fraction.
For the diagram in Fig.~\ref{fig:diags1:b} these curves are also presented as a shading matrix in Fig.~\ref{fig:analysis:b}.
The transition between the liquid and rotator crystal phase is clearly visible, the other transition less so.
Regarding the order of the transitions, we assume that they have to be of first order as continuous symmetries are broken at each of them.

\section{Densities corresponding to the graphs in Fig.~\ref{fig:diags2}}

\begin{figure*}
  \centering
  \subfloat[]{{\includegraphics[width=.66\columnwidth]{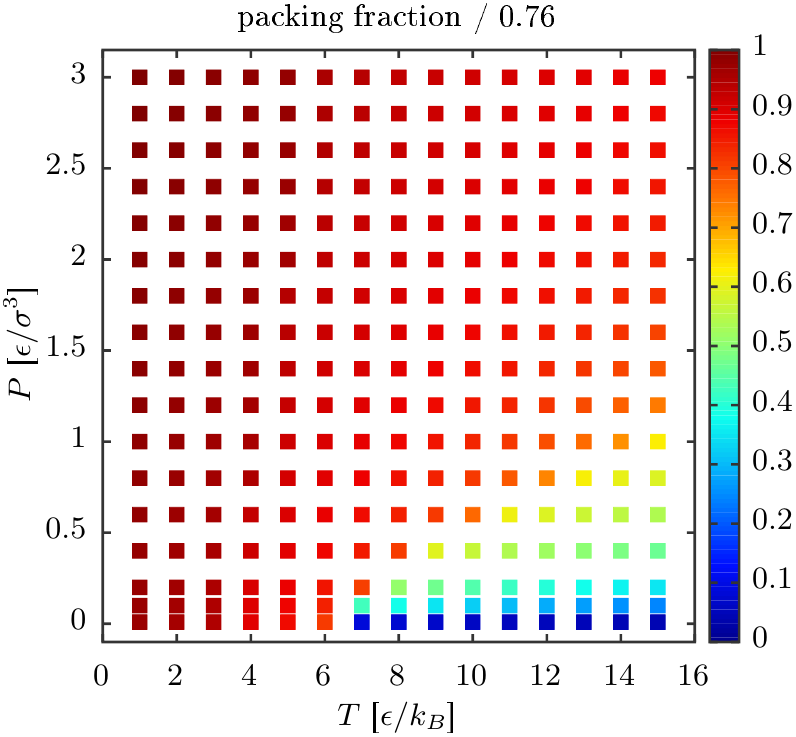}}\label{fig:pf:a}}
  \subfloat[]{{\includegraphics[width=.66\columnwidth]{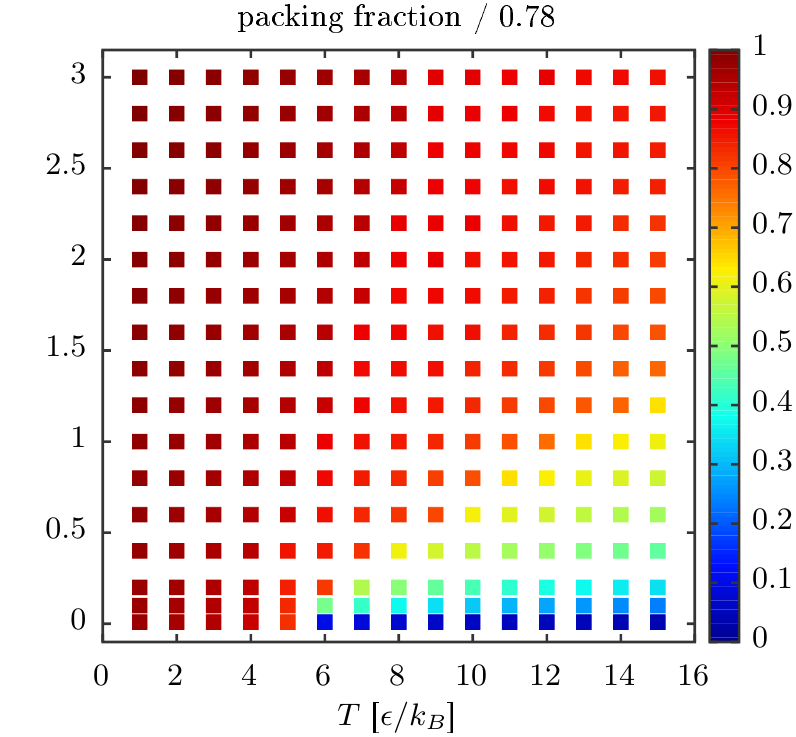}}\label{fig:pf:b}}
  \subfloat[]{{\includegraphics[width=.66\columnwidth]{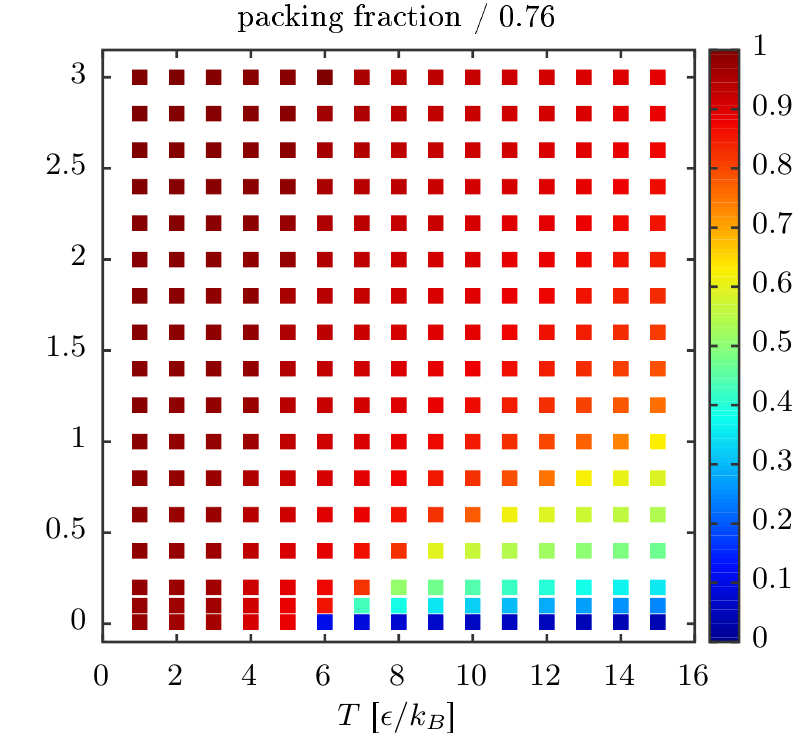}}\label{fig:pf:c}}
  \caption{Packing fractions corresponding to diagrams in Figs.~\ref{fig:diags2:a}, \ref{fig:diags2:b} and \ref{fig:diags2:c}.}
  \label{fig:pfs}
\end{figure*}

In Fig.~\ref{fig:pfs} (also in Fig.~\ref{fig:eos2}) we show the effective packing fractions of the configurations obtained in the diagrams shown in Fig.~\ref{fig:diags2}. Note that these are the densities reached in the entire simulations cell, not in the unit cell of the crystal, i.e.~if defects formed and were not cured during the simulation run, their volumes are included.

\section{Temperature dependence of the orientational correlations}

\begin{figure*}
  \centering
  \includegraphics[width=\columnwidth]{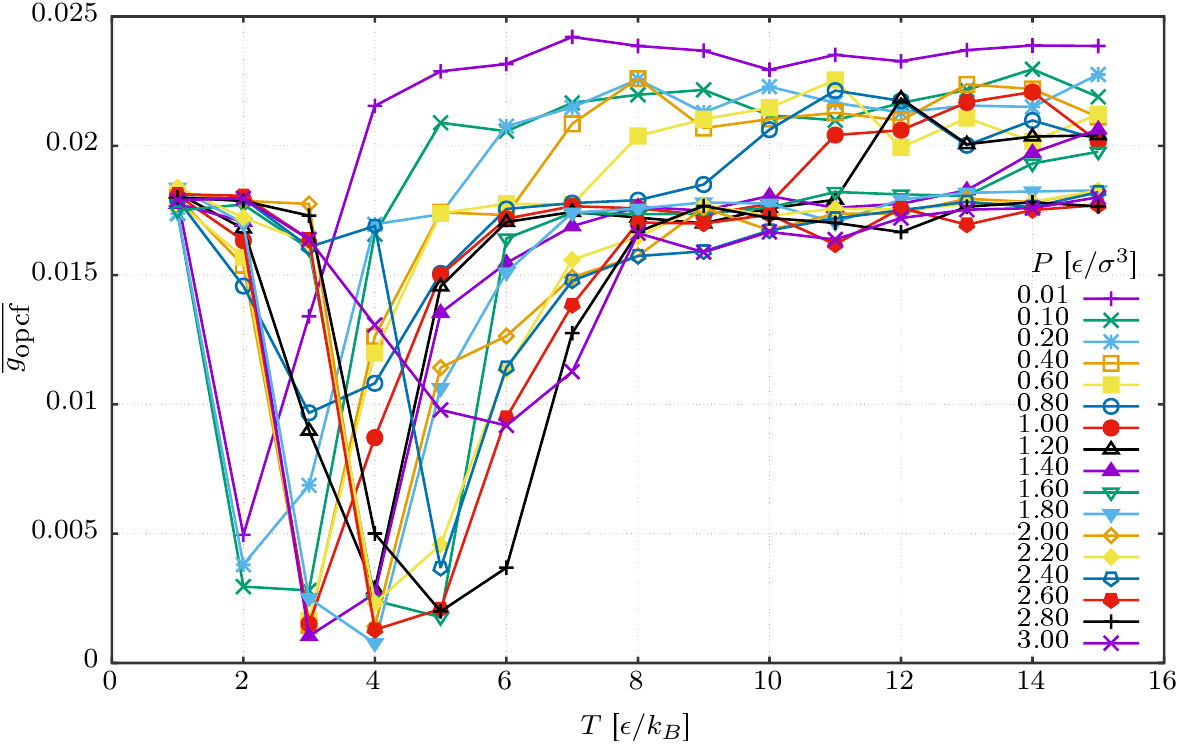}
  \caption{Average value of $g_\text{opcf}$ as a function of temperature at different pressures.}
  \label{fig:opcf2}
\end{figure*}

A more general comparison of the temperature dependence of average value of OPCF for the configurations obtained in Fig.~\ref{fig:diags1:b} is shown in Fig.~\ref{fig:opcf2}.
Large values are obtained for unequilibrated systems at very low temperatures, but from the crystalline configurations on we observe that the orientational order decreases with increasing temperatures.

\end{document}